\documentclass[preprint]{elsarticle}
\usepackage[utf8]{inputenc}
\usepackage{vmargin,tikz}
\usetikzlibrary{calc}

\usepackage{amssymb}
\usepackage{amsthm,amsmath}

\usepackage{caption}
\usepackage{subcaption}
\usepackage{subfloat}
\usepackage{graphicx}
\usepackage{epstopdf}
\usepackage{color}
\usepackage{bibentry}

\newcommand{\br}{{\bf r}}
\newcommand{\cT}{{\cal T}}
\newcommand{\cK}{{\cal K}}

\newcommand{\normal}{\hat{n}(\br)}

\title{Subdivision based Isogeometric Analysis technique for Electric Field Integral Equations for Simply Connected Structures}

\author[ece]{Jie~Li\corref{cor1}}
\ead{ jieli@egr.msu.edu}

\author[ece]{Daniel Dault}
\ead{ daultdan@msu.edu}

\author[CSE]{Beibei Liu}
\ead{liubeibe@msu.edu}

\author[CSE]{Yiying Tong}
\ead{ytong@msu.edu}

\author[ece,phy]{Balasubramaniam~Shanker\corref{cor2}}
\ead{ bshanker@egr.msu.edu}

\cortext[cor1]{Corresponding author}
\cortext[cor2]{Principal corresponding author}

\address[ece]{Department of Electrical and Computer Engineering\\
Michigan Sttae University, East Lansing, MI 48824}

\address[phy]{Department of Physics and Astronomy\\
Michigan State University, East Lansing, MI 48824}

\address[CSE]{Department of Computer Science and Engineering\\
Michigan State University, East Lansing, MI 48824}

\begin{document}

\begin{abstract}
The analysis of electromagnetic scattering has long been performed on a discrete representation of the geometry. This representation is typically continuous but {\em not} differentiable. The need to define physical quantities on this geometric representation has led to development of sets of basis functions that need to satisfy constraints at the boundaries of the elements/tesselations (viz., continuity of normal or tangential components across element boundaries). For electromagnetics, these result in either curl/div-conforming basis sets. The geometric representation used for analysis is in stark contrast with that used for design, wherein the surface representation is higher order differentiable. Using this representation for {\em both} geometry and physics on geometry has several advantages, and is eludicated in Hughes et al., Isogeometric analysis: CAD, finite elements, NURBS, exact geometry and mesh refinement, Computer Methods in Applied Mechanics and Engineering 194 (39-41) (2005). Until now, a bulk 
of the literature on isogeometric methods have been limited to solid mechanics, with some effort to create NURBS based basis functions for electromagnetic 
analysis. In this paper, we present the first complete isogeometry solution methodology for the electric field integral equation as applied to simply connected structures. This paper systematically proceeds through surface representation using subdivision,  definition of vector basis functions on this surface, to fidelity in the solution of integral equations. We also present techniques to stabilize the solution at low frequencies, and impose a Calder\'{o}n preconditioner. Several results presented serve to validate the proposed approach as well as demonstrate some of its capabilities. 

\end{abstract}


\maketitle

\section{Introduction}

Computational methods have become the mainstay of scientific investigation in numerous disciplines, and electromagnetics is no exception. Research in both integral equation and differential equation based methods has grown by leaps and bounds over the past few decades. This period has witnessed development of both higher order basis functions \cite{Hellicar2008,Rieben2005,Jorgensen2004,Graglia1997,Cai2001}, and higher order representations of geometry (at least locally) \cite{Jorgensen2004,Graglia1997,Sevilla2011} amongst many other equally important advances. These approaches have been applied to a wide range of realistic problems spanning several wavelengths. 

However, despite advances in these areas, there is a fundamental disconnect between the geometry processing and analysis based on this geometry. Traditional analysis proceeds by defining a discrete representation of the geometry typically comprising piecewise continuous tessellations. Ironically, this discrete representation of the geometry is obtained using software or a computer aided design (CAD) tool that contains a higher order differentiable representation of the geometry. As eloquently elucidated in \cite{Hughes2005}, the rationale for this disconnect can be attributed to the different periods in time that CAD tools and analysis tools developed. As the latter is older, the computational foundation is older as well. As a result, one is left with awkward communication with the CAD software for refining and remeshing. This is especially true insofar as accuracy is concerned; lack of higher order continuity in geometry can cause artifacts if the underlying  spaces for field representations are not 
properly 
defined. Indeed, the need to define div/curl conforming spaces on  tessellations that are only $C_0$ led to development of novel basis sets that meet this criterion \cite{Nedelec1980}.  An alternate approach that has recently been espoused is isogeometric analysis (IGA). In this approach, the basis functions used to represent the geometry are the same as those used to represent the  physics on this geometry. As a result, the features of geometry representation such as higher order continuity, adaptivity, etc., carry over to function representation as well. Since the appearance of this approach in 2005 \cite{Hughes2005}, it has been applied to a number of applications that range from structure mechanics \cite{Cottrell2006} to fluid-structure interactions (FSI) \cite{Bazilevs2012} to contact problems \cite{DeLorenzis2015} to flow \cite{Nordanger2015} to shell analysis \cite{Kang2015,Bouclier2015} to acoustics \cite{Simpson2014} and electromagnetics \cite{Buffa2010}. In addition to analysis techniques, the 
power of 
IGA has been harnessed for design-through-analysis 
phase in  several practical applications \cite{Kostas2014,Kuru2014,Kiendl2014}.

Next, we briefly review some of the existing methods. Most CAD tools use bi/tri-variate spline based patches/solids like those based upon Bezier, B-splines, and non-uniform rational B-splines (NURBS). As a result, these basis functions are the most often used as IGA basis, with the most popular being NURBS. The latter choice is determined by the fact that NURBS is the industry standard for modern CAD systems. Properties such as non-negativity and the fact that it provides a partition of unity make it an excellent candidate for defining function spaces. Finite element methods based on NURBS basis functions that exhibit $h$- and $p$-adaptivity have been demonstrated \cite{Hughes2005}. Unfortunately, the challenge with using NURBS arises from the fact that the resulting shapes are topologically either a disk, a tube or a torus. As a result, stitching together these patches can result in surfaces that are not 
watertight. 
These 
complexities are 
exacerbated when the object being meshed is topologically complex or has multiple scales \cite{DeRose1998,Bazilevs2010}. Two other geometry processing methodologies  gaining popularity for handling shapes that are complex are T-splines and subdivision surface. The former, an extension to NURBS, can handle T-junctions and hence and greatly reduce the number of the control points in the control mesh. T-splines, especially analysis-ready T-splines,  comprise a good candidate for constructing isogeometric analysis. More detailed work on T-splines and its application in IGA can be found in \cite{Bazilevs2010,Doerfel2010} and references therein. As opposed to T-splines, subdivision surfaces have played a significant role in the computer animation industry. Among its many advantages are the ease with which one can represent complex topologies, scalability, inherently multiresolution features, efficiency and ease of implementation. Furthermore, it converges to a smooth limit surface that is $C^2$ almost everywhere 
except at 
isolated points where it is $C^1$ continuous. There are 
several subdivision schemes; Loop, Catmull-Clark, Doo-Sabin to name a few. Generally speaking, all of the three of these schemes alluded above can be used to construct IGA method. To date, isogeometric analysis based on subdivision surface is less well studied. Some work on IGA based on Catmull-Clark can  be found in \cite{Barendrecht2013}, where IGA is used to solve PDEs defined on a surface.  

While the literature on IGA for differential equations is reasonably widespread across multiple fields, IGA for integral equations is still at a nascent stage. As a result, it has recently become the focus of significant attention. Recently, two dimensional isogeometric boundary element method (IGBEM) was proposed \cite{Simpson2012} to study elastostatic problem with NURBS interpolating basis to represent the geometry, displacement and tractions. In \cite{Scott2013}, IGBEM based on unstructured T-splines was developed for a three dimensional linear elastostatic problem. This approach was extended to address IEs associated with hydrodynamic interactions \cite{Kostas2014}. Likewise, IGBEM methods have been developed to study acoustic scattering from rigid bodies \cite{Simpson2014} as well as two-dimensional electromagnetic analysis \cite{Vazquez2012}. To our knowledge, IGA has not been used for solution to three dimensional IEs associated with vector electromagnetic fields, and this serves to motivate this 
paper.

The focus of this paper will be on the construction of a well formulated low-frequency stable IGA solver for the electric field integral equation (EFIE) that is based on subdivision surfaces specifically, the Loop subdivision scheme. As will be evident, the choice of Loop subdivision scheme is only incidental; the presented method can be applied to most subdivision surface description. In developing a solver that is robust, several challenges need to be addressed; these range from definition of basis functions that correctly map the trace of fields on the surface to formulations that render the resulting system frequency stable to formulation of effective preconditioners. To set the stage for introduction of this formulation, we will assume that the surfaces are simply connected and have $C^2$ smoothness almost everywhere (extension to surfaces with sharp edges and corners is under development and will form the basis of a subsequent paper). As will be evident, the assumption of sufficient ``local'' 
smoothness permits significant freedom in terms of defining function spaces. Thus, the principal contribution of this work is four-fold: We will 
\begin{itemize}
	\item present construction of a basis to correctly represent the trace fields on surface with genus equal to zero,
	\item demonstrate convergence of EFIE-IGA solver for canonical geometries as well as present applications for complex targets,
	\item demonstrate stabilization of EFIE-IGA solvers using the proposed basis sets together with frequency scaling,
	\item demonstrate construction of the Calder\'{o}n preconditioner using the proposed basis sets (without the need for auxiliary barycentric meshes),  
	\item and, present scattering data from multiscale obstacles. 
\end{itemize}

The paper is organized as follows: In Section \ref{sec:prelim} outlines the problem, whereas Section \ref{sec:subd} presents a brief summary of extant literature on subdivision surfaces. Section \ref{sec:em} details the crux of this paper: (i) defines basis functions on the subdivision surface, (ii) develops methods to address low frequency breakdown and multiplicative Calder\'{o}n preconditioners for the EFIE. Section \ref{sec:res} presents several numerical results that verify the accuracy and efficacy of this approach. Finally, Section \ref{sec:summary} summarizes the contribution of the work as well as outlines directions for future research. 

\section{Integral equations for Electromagnetic Scattering \label{sec:prelim}}

The model problem for analysis  will be a perfect electrically conducting object that occupies a volume $\Omega$ whose surface is denoted by $\partial \Omega$. It is assumed that this surface is equipped with an outward pointing normal denoted by ${\hat n}(\br)$.  The region external to this volume ($\mathbb{R}^3 \backslash \Omega$) is occupied by free space. It is assumed that a plane wave characterized by $\{{\bf E}^i ({\bf r}), {\bf H}^i ({\bf r} ) \}$ is incident upon this object. The scattered field for $\br \in \mathbb{R}^3\backslash \Omega$ can be obtained using equivalence theorems as follows: 
\begin{subequations}
\begin{equation}
\begin{split}
\normal \times {\bf E}^s (\br) & = \cT \circ {\bf J} (\br) \\
\normal \times {\bf H}^s (\br) & = \cK \circ {\bf J} (\br)
\end{split}
\end{equation}
where
\begin{equation}
\begin{split}
\cT \circ {\bf J} (\br) & \doteq  - j \omega \mu_0 \normal \times \int_{\partial \Omega} d\br'\left [ {\cal I} + \frac{1}{\kappa^2} \nabla \nabla \right ]g(\br,\br') \cdot {\bf J}(\br')\\
\cK \circ {\bf J} (\br) & \doteq \normal \times \nabla \times \int_{\partial \Omega} d\br' g(\br,\br') {\bf J}(\br)
\end{split}
\end{equation}
\end{subequations}
where $g (\br,\br') = \exp \left [  - j \kappa |\br - \br'| \right ]/(4 \pi |\br - \br'| )$, $\kappa$ is the wave number in free space, ${\bf J}(\br)$ is the equivalent current that is induced on surface, and ${\cal I}$ is the idempotent. In the above expression and in everything that follows, we have implicitly assumed (and suppressed) an $\exp \left [j \omega t \right ]$ time dependence. Using the above equations, one may prescribe the requisite  integral equations as 
\begin{subequations}
\begin{equation}\label{eq:EFIE}
\textbf{EFIE:~~~}\normal \times \normal \times \left ( {\bf E}^i (\br) + {\bf E}^s (\br) \right )= 0~~\forall \br \in \partial \Omega
\end{equation}
\begin{equation}\label{eq:MFIE}
\text{MFIE:~~~}\normal \times \left ( {\bf H}^i (\br) + {\bf H}^s (\br) \right )= 0 ~~\forall \br \in \partial \Omega^-
\end{equation}
\end{subequations}
that are known as the electric/magnetic field integral equations (EFIE/MFIE). In \eqref{eq:MFIE}, $\partial \Omega^-$ denotes a surface that is conformal to but just inside $\partial \Omega$.  These equations do not have unique solutions at the so-called irregular frequencies, but may be combined to yield the combined field integral equation \cite{Mautz1978} that gives unique solution at all frequencies. It should be noted that while these are the most popular formulations that are used in practice, they are not the only ones. Other formulations such as the combined source integral equation \cite{Mautz1979}, augmented EFIE, augmented MFIE \cite{Yaghjian1981} and the charge current integral equation \cite{Taskinen2005} exist and have seen recent development. However, the bulk of this paper's focus will be the EFIE and its discretization. The choice is largely motivated by the numerous challenges that exists in solving these equations, both at the mid and low frequency regimes. To this end, in what follows, we 
will develop basis functions that are well formulated and present modifications to the formulation to account for both low frequency behavior and to impose Calder\'{o}n preconditioners. 

Methods for solving these equations follow the standard prescriptions: (i) represent the surface of the scatterer using some tessellation, (ii) define basis function on this approximation to the geometry, (iii) demonstrate desirable properties of these basis sets, and (iv) validate solutions to integral equations solved using this procedure. As elucidated earlier, as opposed to classical tessellation, we use subdivision surfaces for geometric representation. This is elucidated next. 

\section{Box Splines and Subdivision Surface \label{sec:subd}}

In what follows, we provide a brief overview of shape description as effected by subdivision surfaces. This section is provided purely for completeness and omits details that can be found in several references, e.g., \cite{Cirak2000,Loop1987,Stam1998}. The traditional workhorse of geometric description is NURBS. As a result, current isogeometric methods have been largely developed using NURBS as basis functions. However, a singular feature stands out that poses to be a bottleneck--while NURBS descriptions are $C^2$  in the interior of a patch, the continuity may be $C^0$  or even worse (not watertight) at the boundary between patches or at intersection curves. As a result, one has to define physical basis functions that are div- or curl-conforming for vector problems. This stands in contrast to subdivision surfaces that are constructed as limit surfaces obtained by subdivision processes. This leads to meshes that are $C^2$ almost everywhere except at isolated {\em points} where they are $C^1$. 

\begin{figure}[!ht]
\centering
\subcaptionbox{\label{fig:irregTes_a}}
		[0.5\linewidth]{\includegraphics[width=0.45\linewidth]{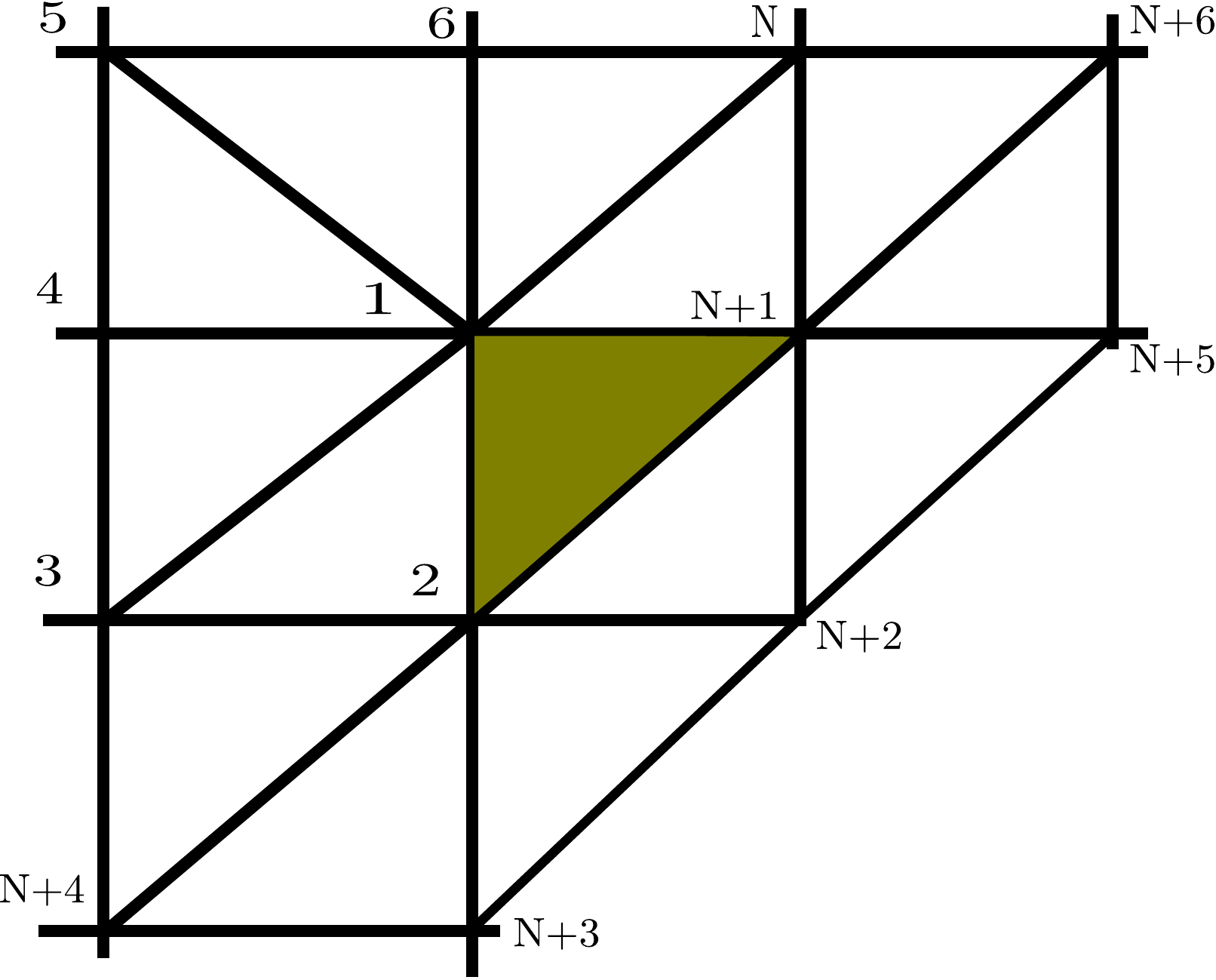}}%
\subcaptionbox{\label{fig:irregTes_b}}
		[0.5\linewidth]{\includegraphics[width=0.45\linewidth]{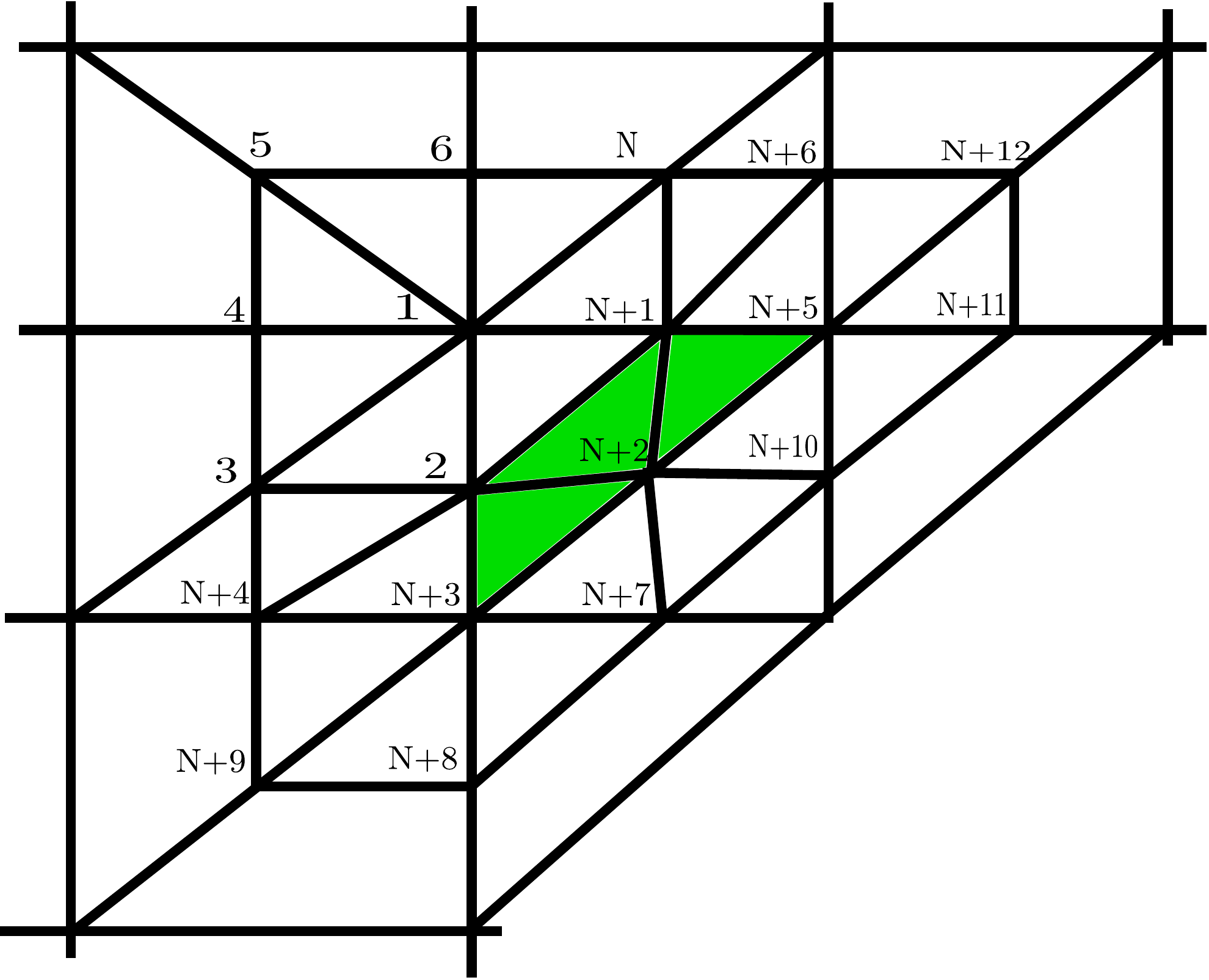}}%
 \caption{ Irregular vertex and triangle (a) An irregular triangle (vertex 1 is valence-7), (b)subdivision once  \cite{Stam1998} }
	\label{fig:irregTes}
\end{figure}

As triangular tessellation is ubiquitous in computational electromagnetics, the choice of the presentation below is based on the Loop subdivision scheme for closed triangular surface meshes. Let $M_0$ denote the primal base mesh that leads to the limit surface. As in any tessellation, $M_0$ comprises a set of vertices $\mathcal{V}$ and a connectivity map. The 1-ring (union of incident triangles) of a vertex produced by the connectivity map can be characterized by the valence of the vertex; specifically, (i) a valence-6 node/vertex is deemed regular and (ii) any other vertex is called \emph{extraordinary} or \emph{irregular}.  A triangle is regular if all its vertices are regular, and irregular otherwise. Fig.~\ref{fig:irregTes} illustrates an irregular triangle and one subdivision around it.

\begin{figure}[ht]
    \centering
    \includegraphics[width=0.8\columnwidth]{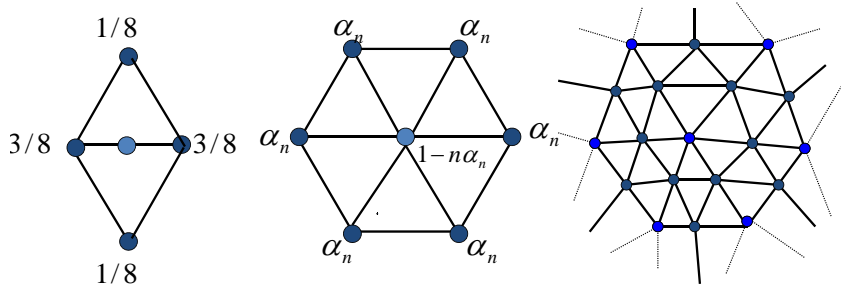} \vspace*{-30mm}
    \caption{Local averaging during refinement.}\vspace*{0mm}
    \label{fig:Loop}
\end{figure}
The Loop subdivision scheme defines a smooth surface approximating positions assigned to each vertex in the base mesh through a procedure (Fig.~\ref{fig:Loop}). Starting from the base mesh, an infinite sequence of meshes is determined through repeatedly refining the $k$-th mesh into the $(k\!+\!1)$-th mesh by dividing each triangle into four subtriangles by inserting one vertex per edge, and adjusting vertex locations. The location of the existing vertex on the refined mesh is an affine combination of its one-ring vertices with weight $\alpha_n=(5/8-(3/8+\cos(2\pi/n)/4)^2)/n,$ and itself with weight $1-n\alpha_n$, where $n$ is the valence. The location of the newly inserted vertex is an affine combination of the end vertices of the edge, each with weight $3/8$, and the other two vertices of the two incident triangles, each with weight $1/8$. The limit of the sequence of piecewise flat surface is the Loop subdivision surface with $C^2$ smoothness almost everywhere, except at the extraordinary vertices, where 
it has $C^1$ smoothness. 
Note, the newly 
inserted vertices will always be regular (valence-6). Thus, we can assume that extraordinary vertices are at least two edges away from each other, otherwise the base mesh is replaced by the mesh obtained after proceeding through the subdivision process once.

The same procedure through repeated refinement with local averaging can be used to define smooth functions on the Loop subdivision surface. In the context of isogeometric analysis, it is perhaps useful to view this process in terms of effective basis functions. Consider the a limit surface $\mathbf{S}(u,v)=\sum \mathbf{x}_i  \xi_i (u,v)$ that is defined by coordinates $\mathbf{x}_i$ of the vertices $V_i$  and $\xi_i(u,v)$ is an effective basis function that is associated with quantities associated with vertex $V_i$, and $(u,v)$ are the pairwise coordinates on a parameterization chart. For instance, $\xi_i(u,v)$ can be  constructed by the subdivision procedure starting with $1$ assigned to $V_i$ and $0$ assigned to all other vertices. 

\begin{figure}[ht]
    \centering
    \includegraphics[width=0.4\columnwidth]{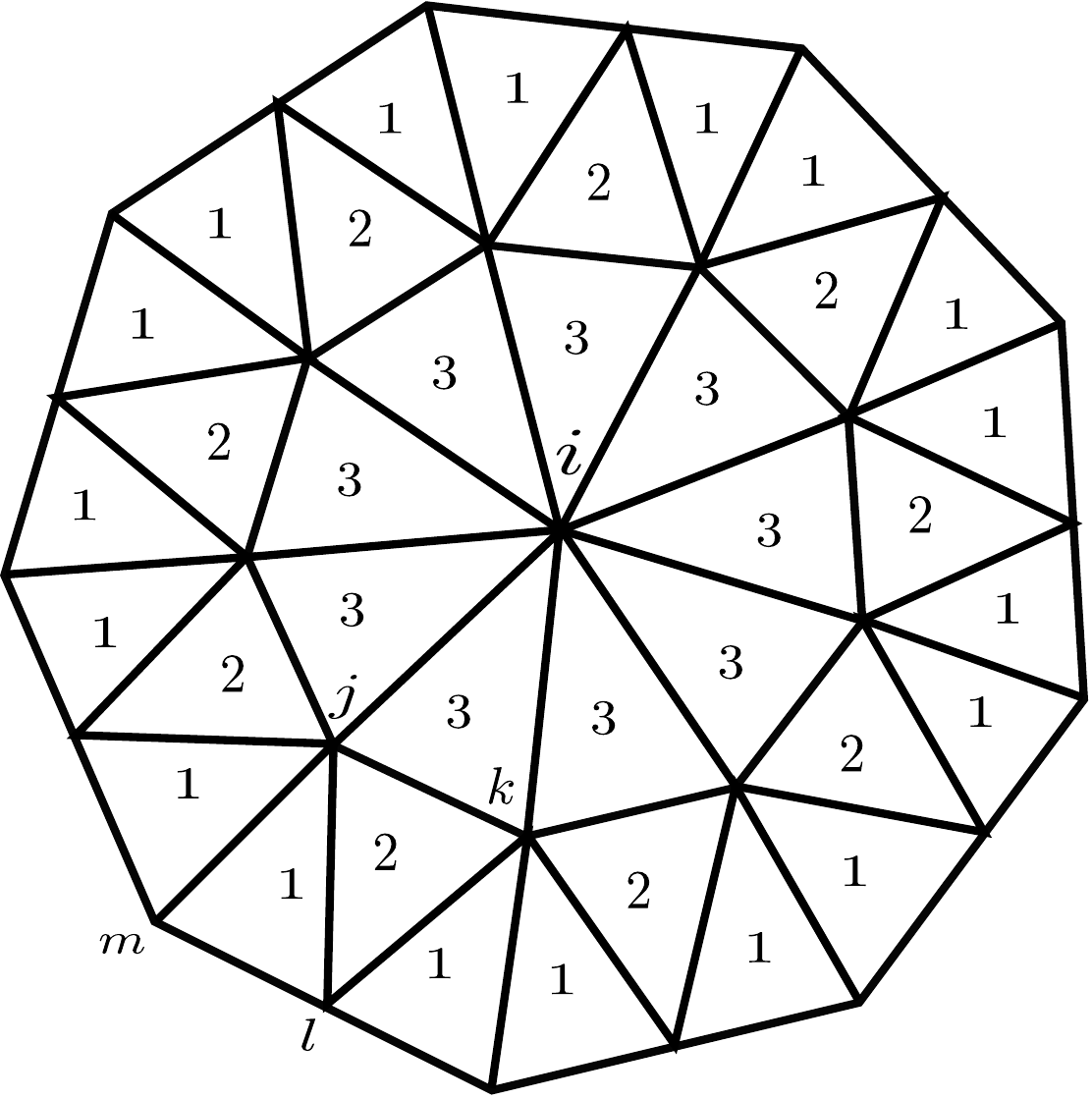} 
    \caption{2-ring of a valence-$k$ vertex ($k=9$ in the example).}
    \label{fig:2-ring}
\end{figure}

Such a basis function $\xi_i(u,v)$ has a compact support, as it is $0$ outside $V_i$'s 2-ring (union of the 1-rings of the 1-ring vertices). Piecewise-linear parameterizations of the base mesh can be used as a parameterization of the subdivision surface. In a parameterization chart $(u,v)$, it can be shown that for any vertex $V_i$, $\xi_i(u,v)$ can be expressed in closed form as the 3-direction quartic box spline $N_i(u,v)$ if $(u,v)$ is located within a regular triangle.  To illustrate this, for triangles incident on $V_i$ (triangles with label 3 in Fig.~\ref{fig:2-ring}), e.g., triangle $T_{ijk}$, denoting the linear basis function (hat function) of $V_i$ by $\varphi_i(u,v)$ and setting $r=\varphi_i$, $s=\varphi_j$ and $t=\varphi_k$ (note that $r+s+t=1$),
\begin{eqnarray*}
N_i(u,v)= 6 r^4+24 r^3 t + 24 r^2 t^2 + 8 r t^3 + t^4 + 24 r^3 s
+ 60 r^2 st + 36 rst^2 \\
+ 6 st^3 + 24 r^2s^2 + 36 rs^2t+12s^2t^2+8rs^3+6 s^3t+s^4.
\end{eqnarray*}
For triangles with 2 vertices in the 1-ring of $V_i$ (triangles with label 2 in Fig.~\ref{fig:2-ring}), e.g., triangle $T_{jlk}$, setting $r=\varphi_j$, $s=\varphi_l$ and $t=\varphi_k$,
\begin{eqnarray*}
N_i(u,v)= r^4+6r^3t+12r^2t^2+6rt^3+t^r+2r^3s+6r^2st+6rst^2+2st^3.
\end{eqnarray*}
For triangles with one vertex in the 1-ring of $V_i$ (triangles with label 1 in Fig.~\ref{fig:2-ring}), e.g., triangle $T_{jml}$, setting $r=\varphi_j$ and $t=\varphi_l$,
\begin{eqnarray*}
N_i(u,v)= r^4+2r^3t.
\end{eqnarray*}

\begin{figure}[ht]
    \centering
    \begin{tikzpicture}[scale=0.7]
\coordinate (O) at (0cm,0cm);
\coordinate (a) at (10cm,0cm);
\coordinate (b) at (0cm,10cm);
\coordinate (oa) at (a);
\coordinate (ob) at (b);
\filldraw[thick,red,opacity=0.3] (O)--(a)--(b)--cycle;
\foreach[count=\xi] \x in {1/2,1/4,1/8,1/16}{%
	\coordinate (oaOld) at (oa);
	\coordinate (obOld) at (ob);
    \coordinate (oa) at ($\x*(a)$);
	\coordinate (ob) at ($\x*(b)$);
	\coordinate (c) at ($(oa)+(ob)$);
	\filldraw[blue,opacity=0.2] (oa)--(c)--(ob)--cycle;
	\pgfmathparse{int (\x*16)};
	\ifnum \pgfmathresult > 1 
		\node (T1) at ($0.333333*(oa)+0.333333*(oaOld)+0.333333*(c)$)  {$\Omega_1^{\xi}$};
		\node (T2) at ($0.333333*(oa)+0.333333*(ob)+0.333333*(c)$)  {$\Omega_2^{\xi}$};
		\node (T3) at ($0.333333*(ob)+0.333333*(obOld)+0.333333*(c)$)  {$\Omega_3^{\xi}$};
	\fi
    }
\end{tikzpicture}
    \caption{The parameter domain within an irregular triangle is recursively partitioned into an infinite number of subdomains \cite{Stam1998}.}
    \label{fig:partitioning}
\end{figure}
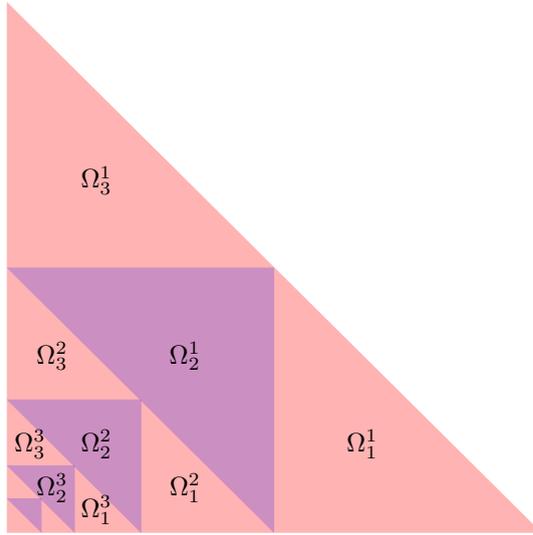

For an irregular triangle $T_{ijk}$ with $V_i$ being extraordinary, the evaluation of $\xi_l$ for $V_l$ in the one-ring of $V_i$, $V_j$ or $V_k$ is modified. If $\varphi_i(u,v)=1$, the evaluation is right at the extraordinary vertex with valence $n$, where it can be shown that $\xi_l(u,v)=1/(\frac{3}{8\alpha_n}+n)$ for $V_l$ in $V_i$'s 1-ring, and $\xi_i(u,v)=1-n/(\frac{3}{8\alpha_n}+n).$ Otherwise, when $2^{-k}<\varphi_i(u,v)\le 2^{-(k-1)}$, the point belongs to $\Omega^k_1$ in Fig.~\ref{fig:partitioning}, if $\varphi_j(u,v)>2^{-k}$, to $\Omega^k_3$ if $\varphi_k(u,v)>2^{-k}$, or to $\Omega^k_2$ otherwise. It means after k-th subdivision, the basis functions ($\xi_i$, $\xi_j$ and $\xi_k$) can be evaluated through linear combinations of box splines with values on the relevant vertices after k-th subdivision, since $\Omega^k_m$ is a regular triangle after $k$-th subdivision. The relevant values for each $\Omega^k_m$ can be stored in a vector $E_{k,m}$ containing $12$ scalars, one for each vertex whose box 
spline basis 
function has the triangle in its support, which can be calculated as
$$E_{k,m}=P_m \bar{A} A^k E_0,$$
where $E_0$ is the initial assignment of the $n+6$ vertices in Fig.~\ref{fig:irregTes_a}, which is $1$ for $V_l$ when evaluating $\xi_l$, and $0$ for any other vertex; $A$ is the subdivision operator matrix, an $(n+6)\times(n+6)$-matrix denoting how the values influencing the irregular triangle in the next level are calculated from the values at the current level (Fig.~\ref{fig:irregTes_b} shows the vertex at the next level); $\bar{A}$ is an extended version of $A$ including how the vertices $n+7$ through $n+12$ of next level is computed form the current level; $P_m$ is the picking matrix selecting the relevant $12$ values for $\Omega^k_m$ within the $n+12$ values. Note that, in practical implementation, eigenanalysis of $A$ is performed so the evaluation involves taking powers of the eigenvalues as an efficient way of handling the calculation of $A^k$. Furthermore, the subdominant eigenvectors provide a means for direct evaluation of derivatives 
even in irregular triangles \cite{Stam1998}. 

Several properties of these basis functions are worth noting as they are critical to represent both the geometry as well as the physics defined on the geometry. Note, some of these properties are available for a NURBS description as well in the interior of the element, but not necessarily across domains. These include the following: 
\begin{enumerate}
	\item compact support,
	\item non-negativity,
	\item convexity preserving,
	\item  and $C^2$ continuity across patch boundaries.
\end{enumerate}
The above properties make this subdivision basis functions a good candidate for modeling physics on both regular and irregular (with the help of subdivision scheme) triangular meshes.

\section{Current representation and field solvers \label{sec:em}}
\subsection{Isogeometric basis sets}
\begin{figure}[!ht]
\centering
\subcaptionbox{\label{fig:bf1}}
		[0.5\linewidth]{\includegraphics[width=0.45\linewidth]{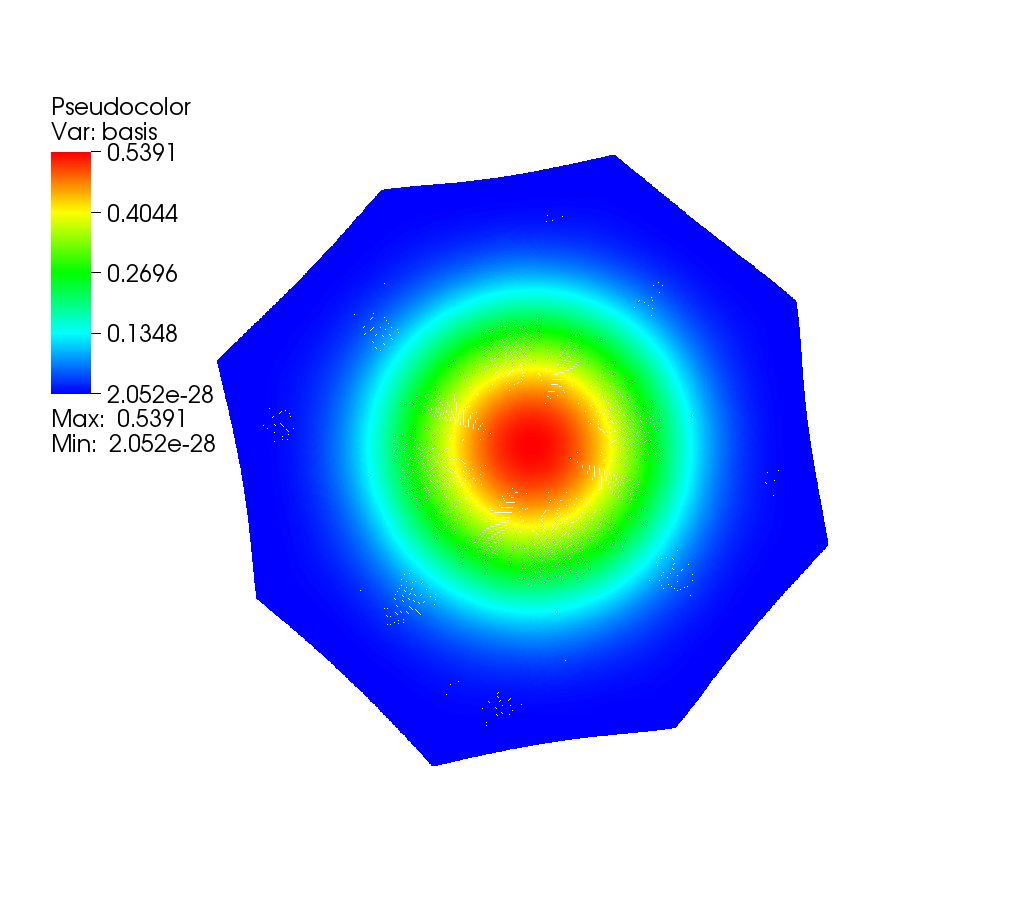}}%
      \subcaptionbox{\label{fig:bf_lap}}
		[0.5\linewidth]{\includegraphics[width=0.45\linewidth]{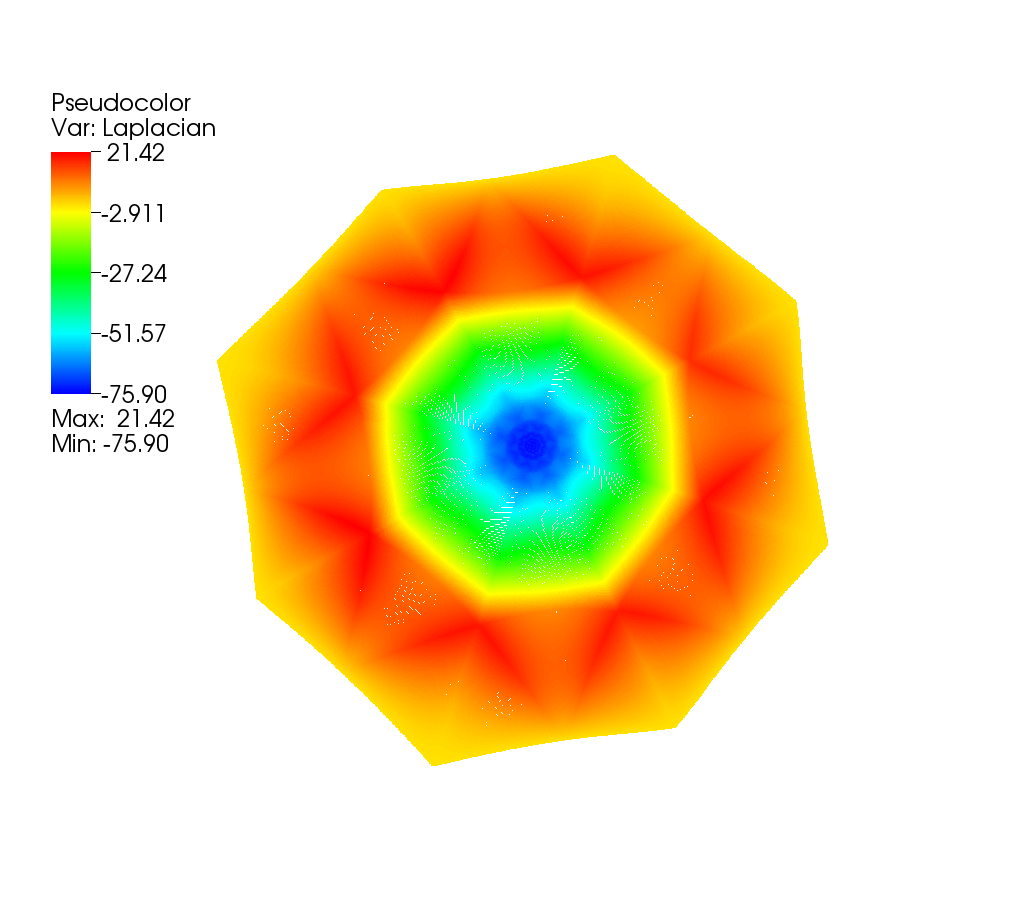}}%
      \caption{ Scalar basis function associated with a vertex with valence of 8. (a) Basis function, (b) Surface Laplacian of the basis function }
	\label{fig:bf}
\end{figure}

Section \ref{sec:subd} developed a framework wherein one dealt with representation of the limit geometry surface. Using the same representation as  presented earlier, we introduce the concept of scalar functions that are defined on the limit surface. As with geometry, the ``limit'' function can be represented in terms of 1-ring control weights of a regular triangle and the box splines as
\begin{equation}
f (\br) = f(\br(u,v)) = \sum_{i=1}^{12} N_i(u,v) w_i; \text{for } \br \in T
\end{equation}
In the above, we have  assumed that the mapping $\br (u,v)$ exists, where $(u,v)$ are the barycentric coordinates of a triangle. In what follows, this assumption will be implicit, and only $\br$ will be used. It can be readily verified that if only one vertex has a weight of unity and other zeros,  one would immediately get a smooth (up to 2nd order continuity globally) scalar function. As a result, one can associate an effective basis function with every vertex such that the limit function 
\begin{equation}\label{eq:scalarBas}
f (\br) = \sum_{n=1}^{N_V} a_n \xi_n (\br)
\end{equation}
where $N_V$ is the number of vertex and $\xi_n (\br)$ is an effective basis function that describes the influence of the scalar quantities associated with a vertex. Fig. \ref{fig:bf1} gives an example of the scalar basis function used for formulating isogeometric analysis on top of subdivision-described surface. The basis function is associated with an irregular vertex of valence 8. To demonstrate 2nd order continuity, the surface Laplacian of the scalar function is plotted in Fig.~\ref{fig:bf_lap}. The behavior of the subdivision basis and its derivatives follow: $\xi_n(\br)\approx O(1)$, $\nabla_s \xi_n(\br)\approx O(1/h)$ and $\nabla^2_s \xi_n (\br) \approx O(1/h^2)$ where $h$ is an approximate dimension of the patch.
\paragraph{Properties of representation} Two salient properties of the scalar basis used for this definition are: (i) as they rely on approximating subdivision they are $C^2$ almost everywhere; and (ii) if $\Omega = \cup_n \Omega_n$, where $\Omega_n$ is the domain associated with  $\xi_n (\br) $, then the function $\xi_n (\br)$ vanishes on $\partial \Omega_n$. In other words, the basis functions $\xi_n (\br) \in C_0^2$ almost everywhere. 

Next, to use this basis set to represent the current, we note that current on any surface can be represented using a Helmholtz decomposition as follows: 
\begin{equation}
 {\bf J} (\br) = \nabla_s \phi (\br)  + \nabla \times ({\hat {\bf n}} \psi (\br) ) + \varpi (\br) 
\end{equation}
where $\varpi (\br) $ is the harmonic field and has a trivial solution ($\varpi (\br) = 0$) for simply-connected structures, and $\phi (\br)$ can be associated with the charge space.  Following the subdivision representation in \eqref{eq:scalarBas}, we will assume that $ \tilde {\phi}(\br)$ and $\tilde{\psi}(\br)$ represent approximations to $\phi (\br)$ and $\psi (\br)$, respectively, and they can be represented in a manner similar to \eqref{eq:scalarBas}; viz., 
\begin{equation}
\begin{split}\label{eq:scalarFlds}
\phi (\br) \approx \tilde{\phi}(\br) & = \sum_n a_{1,n} \xi_n (\br)\\
\psi (\br) \approx \tilde{\psi}(\br) & = \sum_n a_{2,n} \xi_n (\br)
\end{split}
\end{equation}
Using \eqref{eq:scalarFlds}, it is now possible to define the approximations to the current as 
\begin{equation}\label{eq:basisFns}
\begin{split}
{\bf J} (\br) \approx \tilde{{\bf J}} (\br) & = \sum_n \left [ a_{1,n} {\bf J}^1_n (\br) + a_{2,n} {\bf J}^2_n (\br) \right ] \\
{\bf J}^1_n & = \nabla_s \xi_n (\br) \\
{\bf J}^2_n & = \hat{{\bf n} } \times \nabla_s \xi_n (\br) 
\end{split}
\end{equation}
The physical interpretation of this representation is akin to a standard subdivision representation of a limit surface; we represent the the ``limit'' current via the ``limit'' scalar functions. We note, that the basis for the currents (and the auxiliary potentials) are represented via operations on the subdivision basis which can be effected numerically rather trivially. In this work, the standard definition of the inner product $\langle {\bf X}, {\bf Y} \rangle = \int_{\Omega} d\br {\bf X} \cdot {\bf Y}$ is used. 

\paragraph{Properties of the basis function} Several properties of the basis functions make this definition appealing. These are as follows: 
\begin{list}{\labelitemi}{\leftmargin=1em \itemsep=0pt}

\item[\bf Continuity] Due to the reliance on functions $\xi_n (\br)$ that are $C_0^2$ almost everywhere, the  resulting basis functions have $C^1$ continuity almost everywhere and $C^0$ continuity at isolated points. This stands in stark contrast with classical Rao-Wilton-Glisson (RWG) or their higher order counterparts that are div-conforming but not $C^0$ at the boundary. 

\item[\bf Orthogonality] The inner product of $\langle {\bf J}^1_n, {\bf J}^2_n \rangle = 0$. The proof for this assertion can be trivially obtained  using Green's theorems together with properties of the surface curl and the fact that $\xi_n (\br) = 0$ for $(\br) \in \partial \Omega$.  Specifically, 
\begin{equation}\label{eq:hd_orth}
\begin{split}
\int_{\Omega_n} d\br {\bf J}^1_n (\br) \cdot {\bf J}^2_n (\br) & = \oint_{\partial \Omega_n} d\br \xi_n (\br) \hat{u} \cdot {\bf J}^2_n (\br) - \int_{\Omega_n} d\br \xi_n (\br) \nabla_s \cdot {\bf J}^2_n (\br) \\
& = 0 
\end{split}
\end{equation}
where ${\hat u}$ is the normal of $\partial \Omega_n$ tangent to the surface.

\item[\bf Mapping] The function ${\bf J}^1_n (\br)$ maps only to $\nabla_s \phi (\br)$ and ${\bf J}^2_n (\br)$ maps only to $\hat{\bf n} \times \nabla_s \psi (\br)$. The proof can be obtained using Green's theorems, properties of the surface curl/divergence, and the fact that $\xi(u,v)$ is a partition of unity. Specifically, 
\begin{equation}
\begin{split}
\int_{\Omega_n} d\br {\bf J}^1_n (\br) \cdot {\bf J} (\br) & = \oint_{\partial \Omega_n} d\br \xi_n (\br) \hat{u} \cdot {\bf J} (\br) - \int_{\Omega_n} d\br \xi_n (\br) \nabla_s \cdot {\bf J} (\br) \\
& = - \int_{\Omega_n} d\br \xi_n (\br) \nabla_s^2 \phi (\br)
\end{split}
\end{equation}
The key take home message is that ${\bf J}^1_n (\br)$ completely represents the charge space and the rest of the current is represented by ${\bf J}^2_n (\br)$. 

\item [\bf Charge neutrality] The basis functions maintain charge neutrality. This is proved using 
\begin{equation}
\begin{split}
\int_\Omega d\br \nabla_s \cdot \tilde{{\bf J}} (\br) & = \sum_n a_{1,n} \int_{\Omega_n} d\br \nabla_s \cdot \nabla_s \xi_n (\br)\\
& = \sum_n a_{1,n} \oint_{\partial \Omega_n} d\br \hat{u}  \cdot \nabla_s \xi_n (\br) \\
& = 0 \text{ as } \nabla_s \xi_n (\br) = 0 ~\forall {\bf r} \in \partial \Omega_n
\end{split}
\end{equation}

\item [\bf Compactness] The definition of basis function is local.

\item [\bf Refinement] The basis functions are subdivision based. As a result, they inherit properties of  subdivision representation including adaptivity. 
\end{list}

These properties ensure a complete representation of the currents on the surface of an simply connected object, and forms a rigorous Helmholtz decomposition of currents on the surface. Again, as opposed to classical basis functions defined on tessellation, Helmholtz decomposition is inherent in  the definition of basis functions where as the common approach is to design ``quasi''-Helmholtz decomposition using weighted sum of basis functions. It should be noted that, in the above representation,  the potential functions are the actual unknown functions and not the currents. As a result, extra steps have to be taken to ensure their uniqueness.  These are elucidated below when using these basis functions within a Galerkin framework. 

\begin{figure}[!ht]
\centering
\subcaptionbox{\label{fig:bf_loop}}
		[0.5\linewidth]{\includegraphics[width=0.45\linewidth]{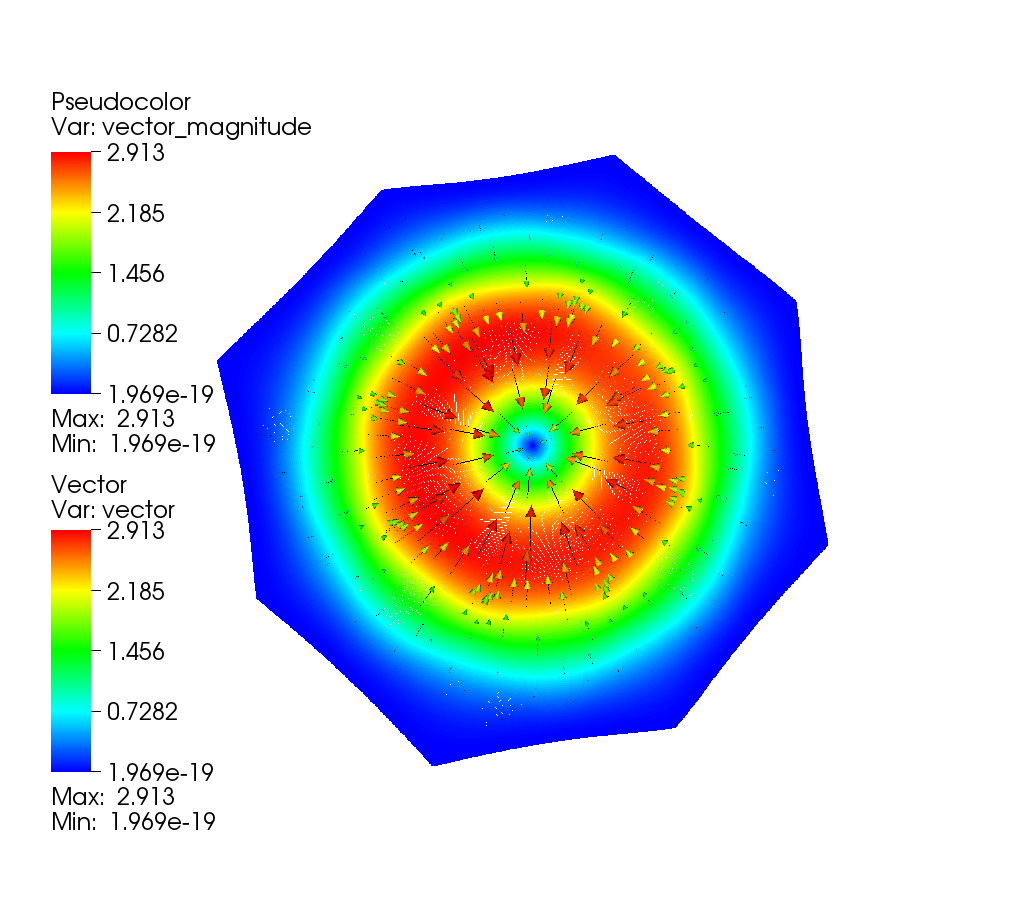}}%
      \subcaptionbox{\label{fig:bf_star}}
		[0.5\linewidth]{\includegraphics[width=0.45\linewidth]{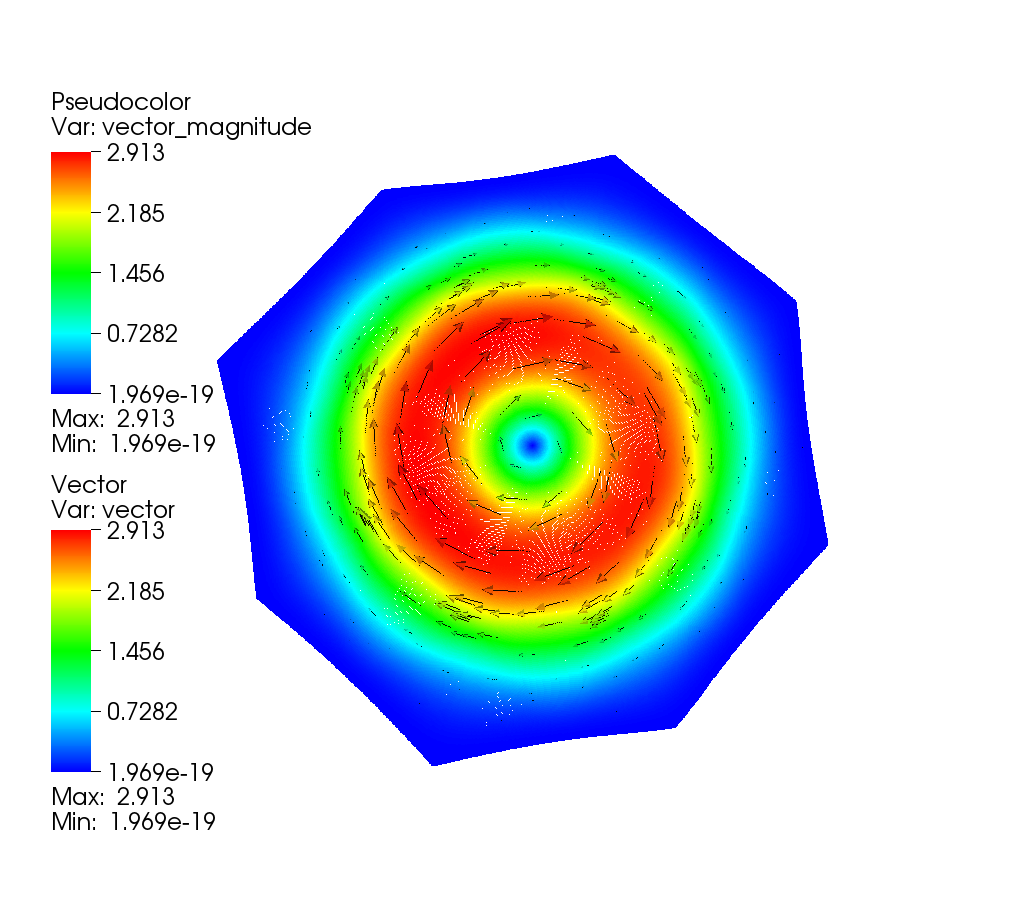}}%
      \caption{ Basis function associated with a vertex with valence of 8. (a) nonsolenoidal type, (b) solenoidal type }
	\label{fig:bf_vector}
\end{figure}

Using these definition of basis functions, each vertex is associated with two degrees of freedom. Fig. \ref{fig:bf_vector} shows the behavior of the two basis functions associated with a node of valence 8. A point to note is that the number of solenoidal and non-solenoidal basis functions are equal to each other. As we will see, this helps creating Calder\'{o}n preconditioners. 

\subsection{Field Solvers}
Given the prescription of basis functions, the discretized version of \eqref{eq:EFIE} can be obtained using Galerkin testing. The resulting matrix equation can be written as 
\begin{subequations}\label{eq:MoM}
\begin{equation}
\left [
\begin{array}{cc}
{\cal Z}^{11} & {\cal Z}^{12} \\
{\cal Z}^{21} & {\cal Z}^{22}
\end{array} \right ] \left\{ \begin{array}{c}
{\cal I}^1\\
{\cal I}^2
\end{array} \right \} = \left \{ 
\begin{array}{c}
{\cal V}^1 \\
{\cal V}^2 
\end{array} \right \}
\end{equation}
where 
\begin{equation}
\begin{split}
{\cal Z}^{lk}_{nm} & = j \omega \mu_0 \int_{\Omega_n} d\br {\bf J}^l_n (\br) \cdot \int_{\Omega_m} d\br' g (\br, \br') {\bf J}^k_m (\br') \\
& - \frac{j\delta_{l1}\delta_{k1}}{\omega \varepsilon_0} \int_{\Omega_n} d\br \nabla_s \cdot {\bf J}^1_n (\br) \int_{\Omega_m} d\br' g (\br, \br') \nabla_s \cdot {\bf J}^1_m (\br')
\end{split}
\end{equation}
\begin{equation}
{\cal I}^k_{n} = a_{k,n}; {\cal V}^k_n = \int_{\Omega_n} d\br {\bf J}_n^k (\br) \cdot {\bf E}^i (\br)
\end{equation}
\end{subequations}
In the above equations, $\delta_{ij}$ denotes a Kronecker's delta. Interesting features of the above expression are apparent; (i) ${\cal Z}^{11}_{nm}$ is the only term that has the charge contribution. As will be shown later, this ``decoupling'' is an essential component for the construction of low frequency stable solvers, and (ii) since the system of equations are constructed using conditions on currents that rely on derivatives of the potential $\tilde{\phi} (\br)$ and $\tilde{\psi}(\br)$, it follows that the potentials are determined upto a constant, and the number of degrees of freedom is less by one for each potential. While one can impose this via different means, we have chosen to essentially constrain the system by choosing $a_{1,N}$ and $a_{2,N}$ to be zero. This implies a trivial change to the system of equations \eqref{eq:MoM}.  The evaluation of the inner products in the above equations is effected via higher order quadrature and Duffy integration rules. While the approach presented thus far is 
valid for all frequencies, it suffers from low frequency breakdown. In what follows, we demonstrate that \eqref{eq:MoM} can be modified trivially to alleviate low-frequency breakdown. Likewise, Calder\'{o}n preconditioners can be easily implemented using this space of basis functions. 

But before we proceed, interesting insight may be gained by examining the inner products that arise using the Galerkin procedure. Matrices arise from testing the electric field, i.e., ${\cal Z}_{nm}^{lk}$ corresponds to measuring the radiated electric field due to ${\bf E}_m (\br) \doteq {\bf E} \left \{ {\bf J}_m^{k} \right \} $ by ${\bf J}_n^l$. If follows that 
\begin{subequations}
\begin{equation}\label{eq:testing_j1}
\begin{split}
\int_{\Omega_n} d\br {\bf J}_n^1 (\br) \cdot {\bf E}_m (\br) & = \int_{\Omega_n} d\br \nabla_s \xi_n (\br) \cdot {\bf E}_m (\br)\\
& = \oint_{\partial \Omega_n} d\br \xi_n (\br) \hat{u} \cdot {\bf E}_m (\br) - \int_{\Omega_n} d\br \xi_n (\br) \nabla_s \cdot {\bf E}_m (\br)  \\
& = - \int_{\Omega_n} d\br \xi_n (\br) \nabla_s \cdot {\bf E}_m (\br)
\end{split}
\end{equation}
\begin{equation}\label{eq:testing_j2}
\begin{split}
\int_{\Omega_n} d\br {\bf J}_n^2 (\br) \cdot {\bf E}_m (\br) & = \int_{\Omega_n} d\br \hat{{\bf n}}\times \nabla_s \xi_n (\br) \cdot {\bf E}_m (\br)\\
& = \int_{\Omega_n} d\br \hat{{\bf n}} \times \nabla \xi_n (\br) \cdot {\bf E}_m (\br)\\
& = \int_{\Omega_n} d\br \hat{{\bf n}} \cdot \left (\nabla \xi_n (\br) \times {\bf E}_m (\br) \right )\\
& = \int_{\Omega_n } d\br \hat{{\bf n}} \cdot \nabla \times \left (\xi_n (\br) {\bf E}_m (\br) \right ) - \int_{\Omega_n} d\br \xi_n (\br) \hat{{\bf n}} \cdot \nabla \times {\bf E}_m (\br)\\
& = \oint_{\partial \Omega_n} d\br \xi_n (\br) {\bf E}_m (\br) \cdot \hat{t} + j\omega \mu_0 \int_{\Omega_n} d\br \xi_n (\br) \hat{{\bf n}} \cdot {\bf H}_m (\br)\\
& =  j\omega \mu_0 \int_{\Omega_n} d\br \xi_n (\br) \hat{{\bf n}} \cdot {\bf H}_m (\br)
\end{split}
\end{equation}
\end{subequations}
In the above equation, $\hat{t}$ is a unit vector tangential to the boundary $\partial \Omega_n$, and ${\bf H}_n (\br)$ is the magnetic field due to ${\bf J}_m^k$. From the above equations, it is apparent that \eqref{eq:testing_j1} tests the tangential component of the electric field. However, \eqref{eq:testing_j2} yields equations that test in normal component of the magnetic field. So the two  basis functions used in the analysis impose tangential continuity of the electric field as well as the normal component of the magnetic field. As a result, using a basis that satisfies the Helmholtz decomposition (and relies on scalar function that are $C^2_0$), results in equations that naturally fit into the framework of the Current-Charge Integral Equations (CCIE) \cite{Taskinen2005}. It should also be noted that while eqns. \eqref{eq:testing_j1} and \eqref{eq:testing_j2} were specified for the scattered field, they are equally  valid for the incident field. Indeed, using the final expression 
in \eqref{eq:testing_j2} to evaluate the 
integrals is more 
accurate and requires fewer quadrature points. 

\subsection{Low-frequency stable EFIE}

It is well known that the EFIE suffers from low-frequency breakdown. The rationale for this breakdown is readily apparent by examining the components of the impedance matrix in \eqref{eq:MoM}. 
Assuming that the average largest linear dimension of the support of the patch is $h$. It follows that the entries corresponding to 
\begin{equation}\label{eq:scale}
{\cal Z}^{lk}_{nm} = {\cal O} (\kappa h^2) + \delta_{l1}\delta_{k1} {\cal O} (1/\kappa)
\end{equation}
These results follow from the fact that the functions $\xi (\br) = {\cal O} (1)$. The above scaling indicates that a portion of the elements associated with source and test  basis functions that are associated with irrotational functions ${\bf J}^1_n (\br)$ scale as ${\cal O}(1/\kappa)$, whereas all others scale as ${\cal O} (\kappa h^2)$. This implies that as $\kappa \longrightarrow 0$, the portion of ${\cal Z}^{11}_{nm}$ that corresponds to the charge contribution dwarfs the rest. This situation is similar to those encountered for by classical Nedelec elements. Here, one takes recourse to loop-star/loop-tree decompositions \cite{wu1995,Vecchi1999,Andriulli2012} that effect an approximate Helmholtz decomposition of the currents. As has been shown, the resulting decompositions contain a portion that is exactly divergence free and one that is approximately curl free. Whereas the support of the divergence-free portion is local, the same is not true of curl-free portion. This is in contrast to the method 
presented here 
wherein both components have 
local support. Using these basis results in the charge being modeled correctly, and a matrix whose scaling looks like \eqref{eq:scale}. Rescaling these equations has been shown to render the solution stable. Following a similar procedure, it can be shown that rescaling both the matrix elements, the coefficients and the right hand side results in 
\begin{equation}
\begin{split}
{\cal Z}^{lk}_{nm} & = j \omega \beta_{lk} \mu_0 \int_{\Omega_n} d\br {\bf J}^l_n (\br) \cdot \int_{\Omega_m} d\br' g (\br, \br') {\bf J}^k_m (\br') \\
& - \frac{j\delta_{l1}\delta_{k1}}{\varepsilon_0} \int_{\Omega_n} d\br \nabla_s \cdot {\bf J}^1_n (\br) \int_{\Omega_m} d\br' g (\br, \br') \nabla_s \cdot {\bf J}^1_m (\br')
\end{split}
\end{equation}
where 
\begin{equation}
\beta_{lk} = \left \{ \begin{array}{cc}
\omega & \text{ if } l = 1, k = 1\\
\omega^{-1} & \text{ if } l = 2, k = 2\\
1 & \text {otherwise}
\end{array} \right . 
\end{equation}
\begin{equation}
{\cal I}^k_{n} = a_{k,n}\omega^{-\delta_{k1}}; {\cal V}^k_n = \omega^{-\delta_{k2}}\int_{\Omega_n} d\br {\bf J}_n^k (\br) \cdot {\bf E}^i (\br)
\end{equation}
As prescribed, these equations achieve the desired stability. As can be trivially shown, these equations decouple  as $\omega \longrightarrow 0$. This is akin to similar prescriptions for classical loop-star/tree algorithms \cite{wu1995,Vecchi1999,Lee2003}. However, a {\em salient} feature of the IGA-basis is that constructing a well behaved system is tantamount to using a {\em diagonal preconditioning} sans constructing the complementary system that is required for a loop-tree/star (Hodge) decomposition. 

It should be noted that Helmholtz decomposition is not the only way to solve low frequency breakdown, and other techniques \cite{Taskinen2005,Qian2010,Epstein2010,Epstein2013,Xiong2013} exist. The isogeometric basis functions presented herein could help construct those methods.

\subsection{Calder\'{o}n preconditioner}

It is well known that the standard EFIE presented in the earlier sections is a first kind equation, and as with all first kind equations, can be ill-conditioned especially when the spatial scales in the problem are widely separated. As has been shown by several others (see Refs. \cite{Hsiao1997,Vipiana2007,Andriulli2008a}, and references therein), the rationale for the ill-conditioning is the spectral separation of the eigenvalues of the operator with increase in discretization density, with a set that clusters around zero and others that cluster around infinity. Given this separation, the resulting matrix systems become rapidly ill-conditioned. The remedy to this problem exploits the Calder\'{o}n identities wherein the EFIE operator (${\cal T} \left \{ \cdot \right \} $) preconditions itself resulting in a second kind integral operator whose eigenvalues accumulate around $-1/4$ \cite{Hsiao1997,Christiansen2002,Andriulli2008}. The challenge in using these identities was the lack of well behaved Gram matrices 
that link the domain and range of the ${\cal T}$ 
operator\cite{Andriulli2008,Yan2010}. For Thomas-Raviart/Rao-Wilton-Glisson basis sets and $C^0$ geometries, 
basis functions developed by Buffa and Christiansen \cite{Buffa2007} have been exploited in a sequence of  papers to thoroughly understand and solve this problem. In what follows, we show that basis functions defined herein result in a well conditioned Gram matrix, thereby permitting a natural discretization of the Calder\'{o}n operator. To begin, we define operators
\begin{subequations}
\begin{equation}\label{eq:CalderonSplit}
{\cal T} \circ {\bf J} (\br) = {\cal T}_a \circ {\bf J} (\br) + {\cal T}_\phi\circ {\bf J} (\br) 
\end{equation}
where 
\begin{equation}
\begin{split}
{\cal T}_a \circ {\bf J} (\br) & \doteq - j \omega \mu_0 \normal \times \int_{\partial \Omega} d\br' g(\br,\br') \cdot {\bf J}(\br')\\
{\cal T}_\phi \circ {\bf J}(\br) & \doteq - j \frac{1}{\omega \varepsilon_0 }\normal \times \int_{\partial \Omega} d\br'  \nabla \nabla g(\br,\br') \cdot {\bf J}(\br')
\end{split}
\end{equation}
\end{subequations}
The Calder\'{o}n projector is defined as ${\cal T}^2 \circ {\bf J} (\br) \doteq {\cal T} \circ {\cal T} \circ {\bf J} (\br)$. It has been shown that 
${\cal T}^2 \circ {\bf J} (\br) = (-1/4 + {\cal  K}^2 ) \circ {\bf J}(\br)$, where ${\cal K}^2 \circ {\bf J}(\br) \doteq  {\cal K} \circ {\cal K} \circ {\bf J}(\br)$ is compact, and the eigenvalues are clustering around $-1/4$. Alternatively, this operator may also be written as  ${\cal T}^2 \circ {\bf J} = \left ( {\cal T}_a \circ {\cal T}_a  + {\cal T}_a \circ {\cal T}_\phi + {\cal T}_\phi \circ {\cal T}_a + {\cal T}_\phi \circ {\cal T}_\phi  \right ) \circ {\bf J}$. Since direct discretization of ${\cal T}^2$ is impossible, typical implementation follows the multiplier approach; i.e., define intermediate mapping from the range space of the ${\cal  T}$ operator to the domain of the ${\cal T}$. This is typically effected via a Gram matrix and has been extensively explored.  Given the usage of Helmholtz decomposition \eqref{eq:basisFns} and the orthogonality \eqref{eq:hd_orth} between the two components, it can be shown that ${\cal T}_\phi \circ {\cal T}_\phi \circ 
{\bf J} = 0$. 
Enforcing the latter condition has been difficult to accomplish in 
traditional discretization and representation schemes. 

The principal advantage of using \eqref{eq:basisFns} is that it enforces an exact global Helmholtz decomposition in terms of a set of local scalar functions that are $C_0^2$ almost everywhere. Recall the following: $\langle {\bf J}^l_n, {\bf J}^k_m \rangle = \gamma_{nm} \delta_{lk}$, where $\gamma_{nm}$ is the result of evaluating the inner product over the support of basis functions $\Omega_n \cap \Omega_m$ and is defined in \eqref{eq:gammaMN}. As a result, the Gram matrix ${\cal G }$ is block diagonal with entries ${\cal G}_{nm}^{lk} = \gamma_{nm} \delta_{lk}$. Given that, for a pair of spaces,  $ ({\bf J}^1, {\bf J}^2)$, it can be shown that the following sequence is satisfied: $\left ( {\bf J}^1, {\bf J}^2 \right) \xrightarrow{{\cal T}} \left ({\bf J}^2, {\bf J}^1 \right ) \xrightarrow{{\cal T}} \left ( {\bf J}^1, {\bf J}^2\right)$. Specifically,  $\left ( {\bf J}^1, {\bf J}^2 \right) \xrightarrow{{\cal T}_\phi} \left ({\bf J}^2, 0\right )$ and  $\left ( {\bf J}^1, {\bf J}^2 \right) \xrightarrow{{\cal T}
_a} \left ({\bf J}^2, {\bf J}^1\right )$. It follows from the above that $\left ( {\bf J}^1, {\bf J}^2 \right) \xrightarrow{{\cal T}_\phi} \left ({\bf J}^2, 0\right ) \xrightarrow{{\cal T}_\phi} \left ( 0,0\right)$. In matrix form, the above sequence can be rewritten as
\begin{equation}
{\cal T} {\cal G}^{-1} {\cal T} = \left[ 
\begin{array}{cc}
{\cal T}_a^{11} & {\cal T}_a^{21}\\
{\cal T}_a^{21} + {\cal T}_\phi ^{21} & {\cal T}_a^{22}
\end{array}\right ] \left[\begin{array}{cc}
{\cal G}^{11} & 0 \\
0 & {\cal G}^{22}
\end{array}
\right ]^{-1} \left [ \begin{array}{cc}
{\cal T}_a^{11} & {\cal T}_a^{21} \\
{\cal T}_a^{21} + {\cal T}_\phi ^{21} & {\cal T}_a^{22}
\end{array}\right ]
\end{equation}
Finally, a critical component in discretizing the ${\cal T}^2$ operator is the Gram matrix. The benefits of using local Helmholtz decomposition becomes apparent in its construction with the off-diagonal blocks being zero and the element of the diagonal blocks being  
\begin{equation}\label{eq:gammaMN}
\gamma_{nm} = \int_{\Omega_n \cap \Omega_m} d\br \nabla_s \xi_n(\br) \cdot \nabla_s \xi_m (\br)
\end{equation}
that are a variational form of the Laplace-Beltrami operator on a scalar function. The resulting system is positive definite leading to well behaved inverse of the Gram system.

\section{Numerical Examples} \label{sec:res}
This section presents several numerical examples to demonstrate the efficacy of the proposed approach to electromagnetic analysis. In order to do so, we shall present data that demonstrates the following; (i) accuracy of the approach to when compared against analytical data;  (ii) examples illustrating low frequency stability, (iii) analysis of multiscale structures to illustrate the efficiency of the Calder\'{o}n preconditioner, and (iv) application to complex targets,. Unless stated otherwise, the data presented compares in radar scattering cross-section computed using the IGA-MoM solver and those computed using either analytical methods (if available) or a validated method of moments code that is based on RWG basis functions. 

\subsection{Accuracy of IGA-MoM} \label{sec:iga_validate}
 
\begin{figure}[!ht]
\centering
\subcaptionbox{\label{fig:cur_sph_real}}
		[0.5\linewidth]{\includegraphics[width=0.3\linewidth]{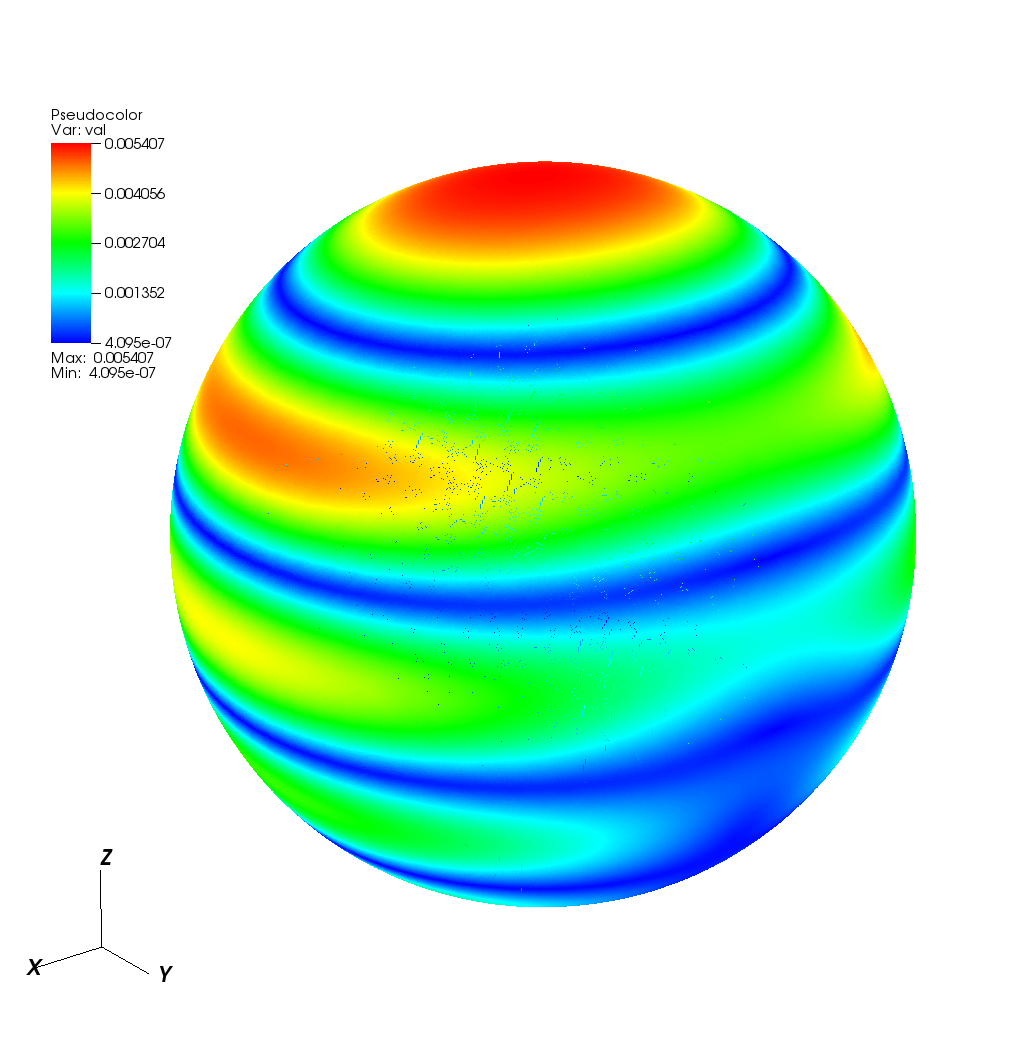}}%
      \subcaptionbox{\label{fig:cur_sph_imag}}
		[0.5\linewidth]{\includegraphics[width=0.3\linewidth]{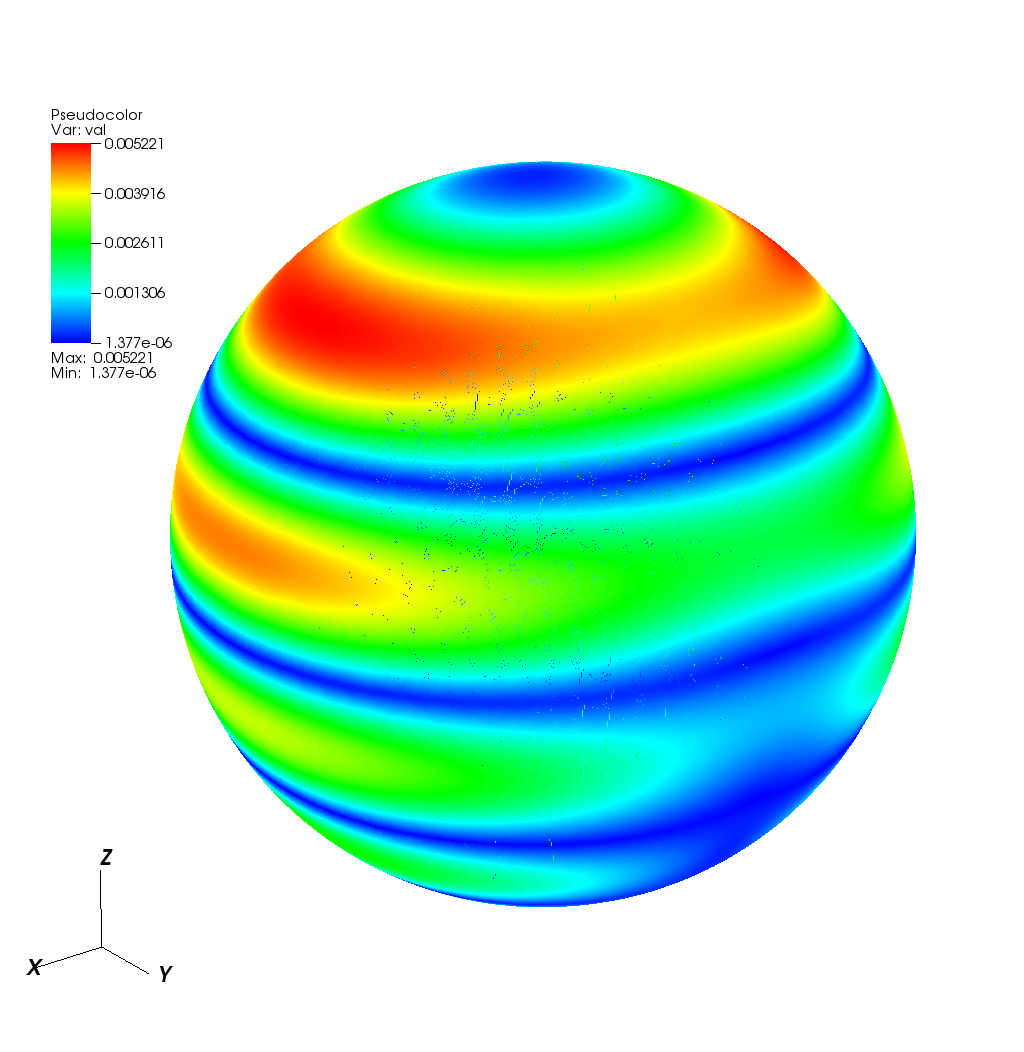}}%
      \caption{ Magnitude of surface currents density on the sphere: (a) real part and (b) imaginary part }
	\label{fig:sphere_currents}
\end{figure}

\begin{figure}[!ht]
\centering
\subcaptionbox{\label{fig:err_real}}
		[0.5\linewidth]{\includegraphics[width=0.45\linewidth]{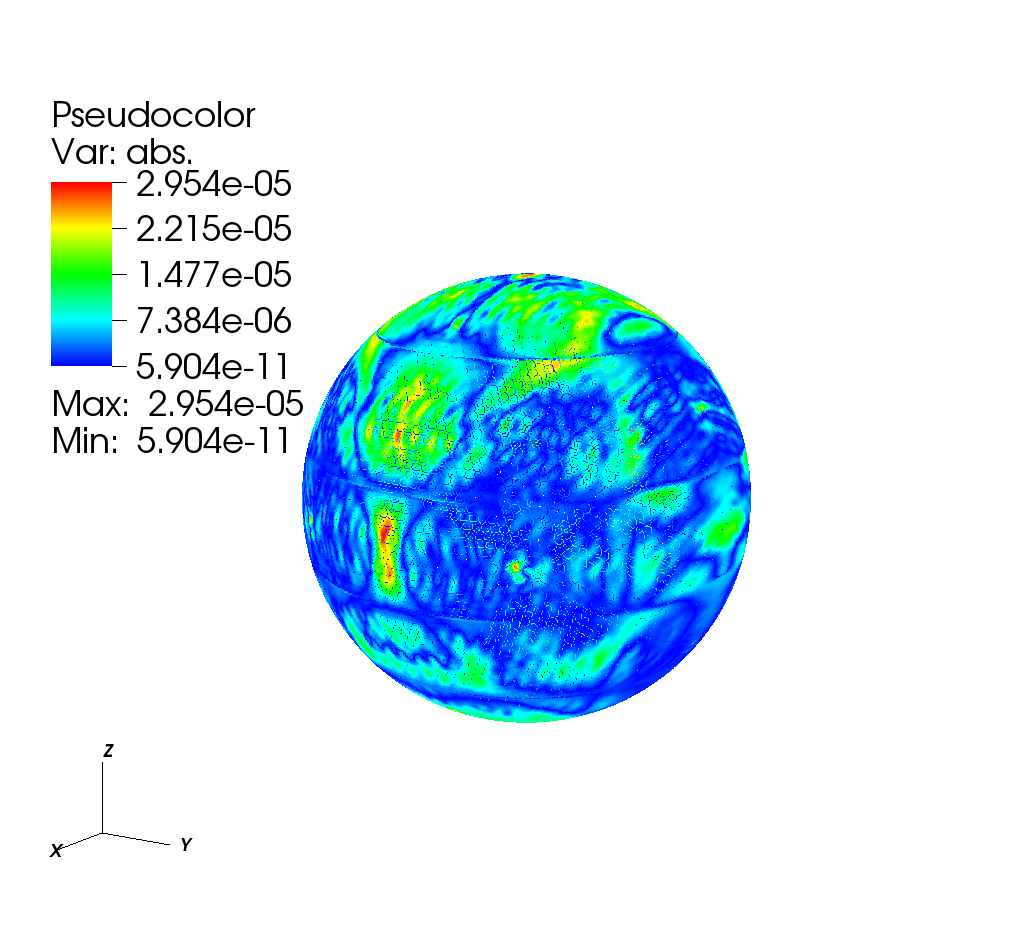}}%
      \subcaptionbox{\label{fig:err_imag}}
		[0.5\linewidth]{\includegraphics[width=0.45\linewidth]{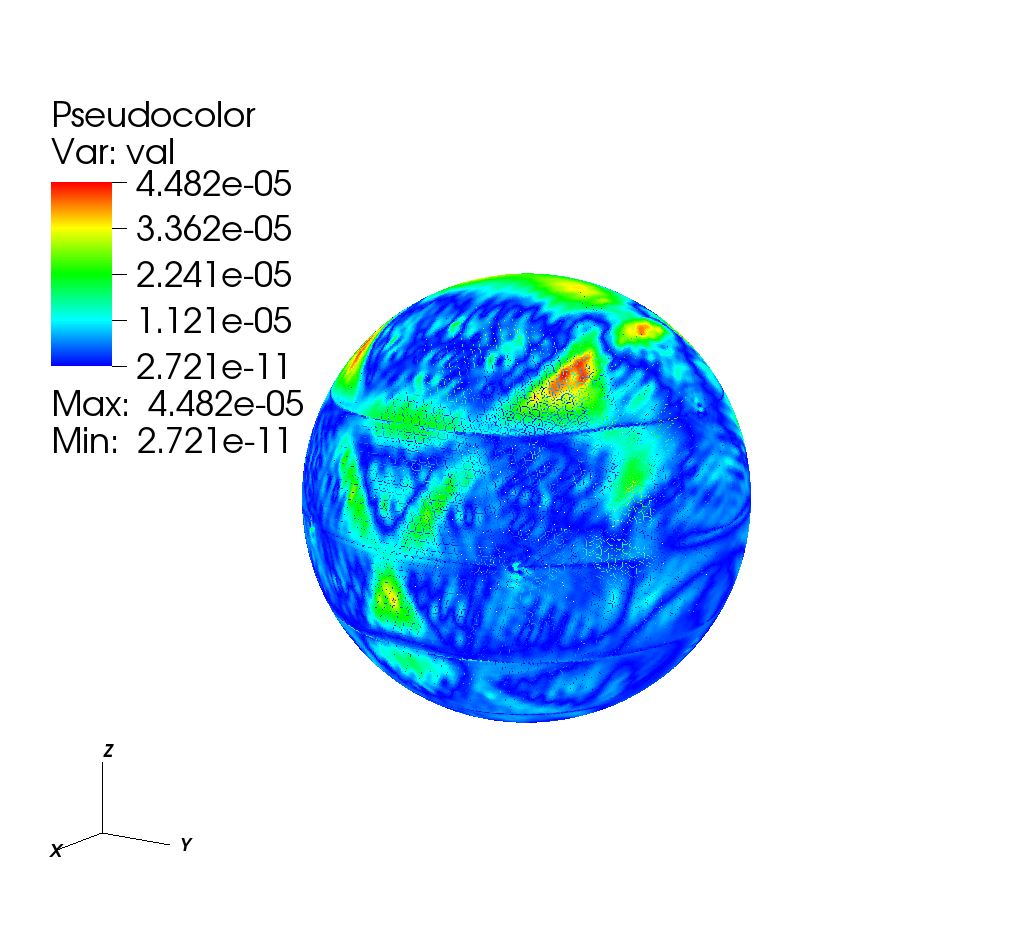}}%
      \caption{ Pointwise relative error (a) real part, (b) imaginary part with 20484 DoFs}
	\label{fig:err_sphere}
\end{figure}

To validate and demonstrate the accuracy of the proposed approach, we consider scattering from a sphere discretized at multiple resolutions. To this end, consider a  sphere with radius of $1\lambda$ that is modeled using an initial control mesh comprising 642 vertices. Starting from this control mesh and performing one and two Loop subdivision results in two other meshes with 2562 and 10242 vertices, respectively. This results in three meshes with $1280$, $5120$, and $20480$ faces. Note, the limit surface for all three mesh densities is identical. As a result, what changes is the support of the basis functions, and therefore, approximations to the discrete operators. In all the data presented below, a field propagating in $\hat{k} = -\hat{z}$ and polarized along $\hat{x}$ axis is incident on the sphere. The number of degrees of freedom (DoF) for IGA-MoM is twice the number of vertices; consequently, the three tests have 1284, 5124, and 20484 DOFs, respectively. Table \ref{tab:tab1} presents errors in 
currents on the 
surface of the sphere (between those obtained using IGA-MoM and Mie series) for these three different discretizations. Likewise, Figs. \ref{fig:sphere_currents} and \ref{fig:err_sphere} present pointwise errors in absolute values of the real and imaginary parts of the current for the case of two subdivisions. As is evident, the accuracy in the current both in the $L_2$ and $L_1$ norms is excellent.

  \begin{table}
\centering
  \caption{ Rel. $L_2$ error in the currents density }
    \begin{tabular}{ c | c c c }  
  \hline
     subdivision times & 0 & 1  & 2  \\ \hline
     average  element size &  0.1511 & 0.0755  & 0.0377     \\ 
    Rel. $L_2$ error &  0.0257 & 0.0067  & 0.0030     \\ 
    \hline
    \end{tabular}
\label{tab:tab1}
\end{table}

\begin{figure}[!ht]
\centering
\includegraphics[width=0.8\linewidth]{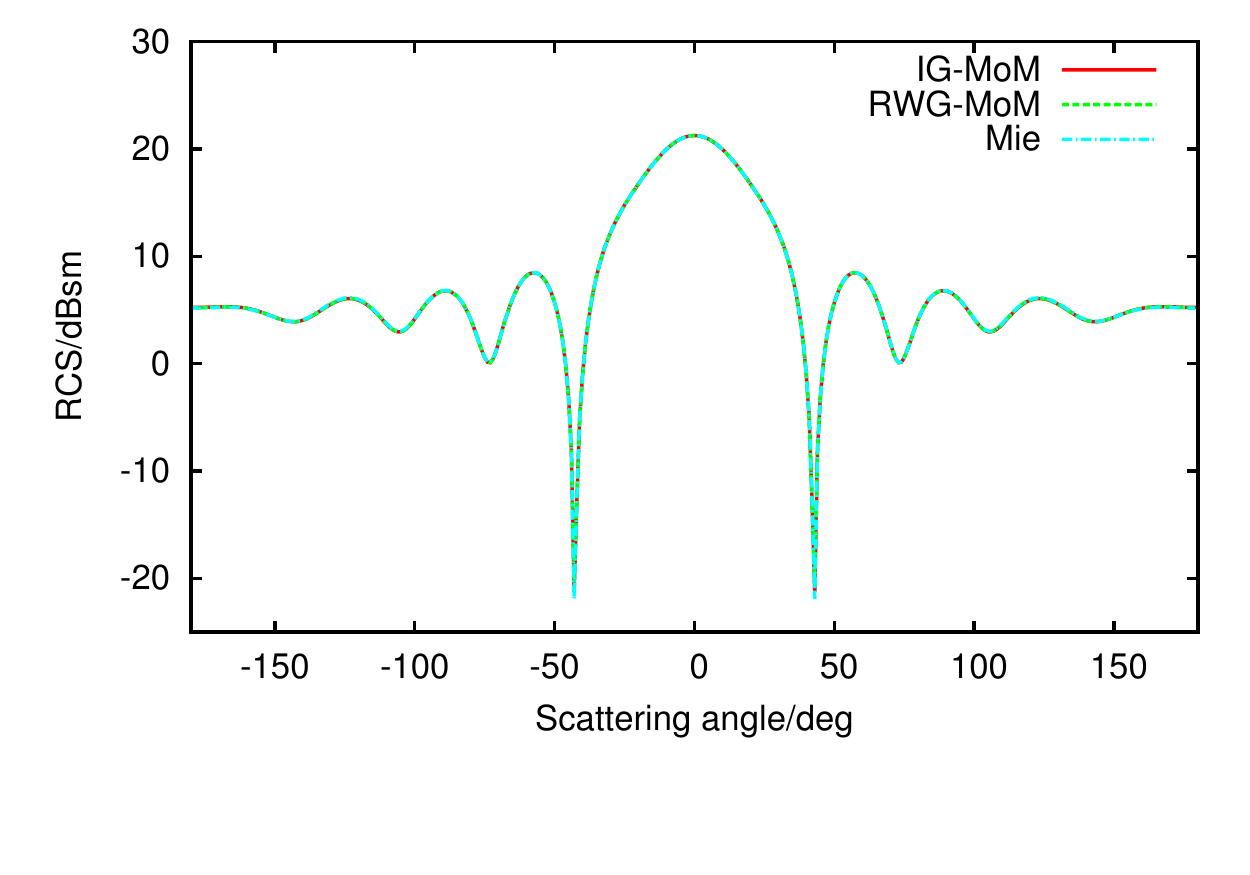}
\caption{Radar cross section of the sphere ($\phi=0$ cut)}
	\label{fig:sphere_rcs}
\end{figure}

 \subsection{Scattering from Structures at regular frequency}


In what follows, we analyze the performance of the proposed approach and compare these with a conventional MoM EFIE solution technique that relies on RWG basis functions. Unless specified otherwise, the metric for comparison is RCS data in the $\phi= 0$ plane.  Furthermore, the geometry of all scatterers that are analyzed in this paper is smooth; extension to non-smooth structures is non-trivial but realizable from a subdivision perspective \cite{Ling2008}. IGA solvers that include in open and sharp surfaces will form the basis of a future work. 

\begin{figure}[!ht]
\centering
\subcaptionbox{\label{fig:cone_real}}
		[0.5\linewidth]{\includegraphics[width=0.45\linewidth]{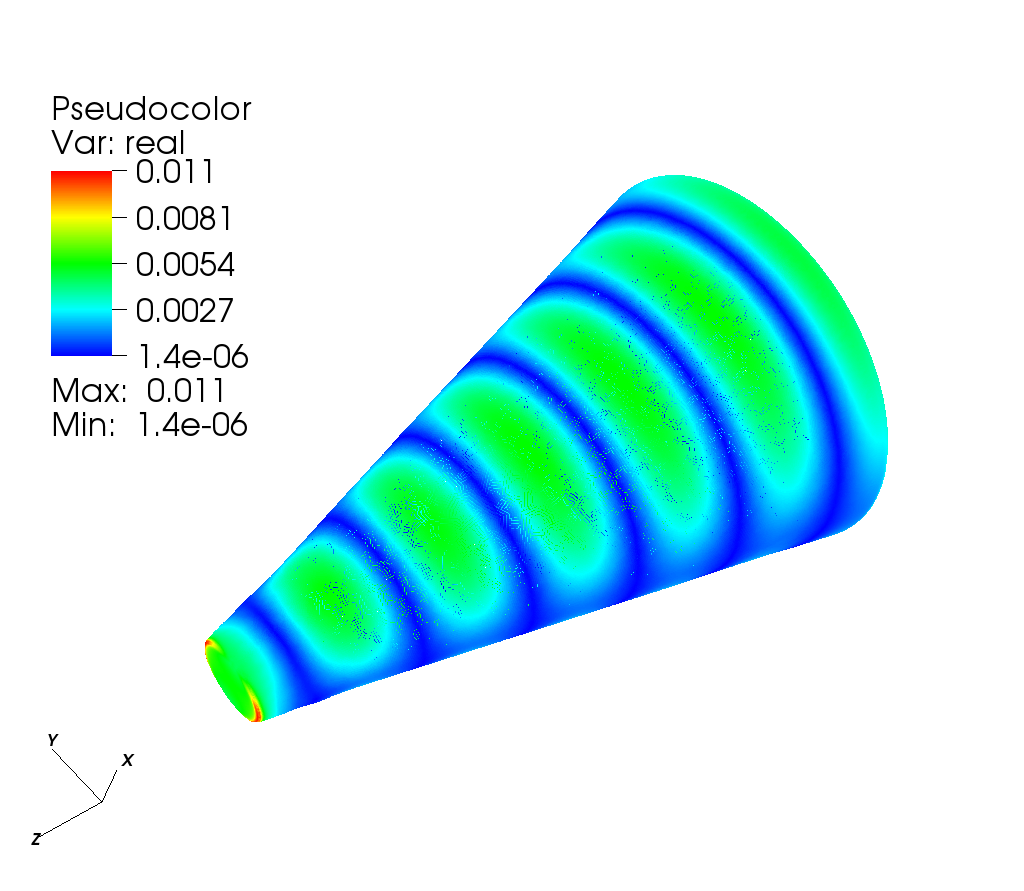}}%
      \subcaptionbox{\label{fig:cone_imag}}
		[0.5\linewidth]{\includegraphics[width=0.45\linewidth]{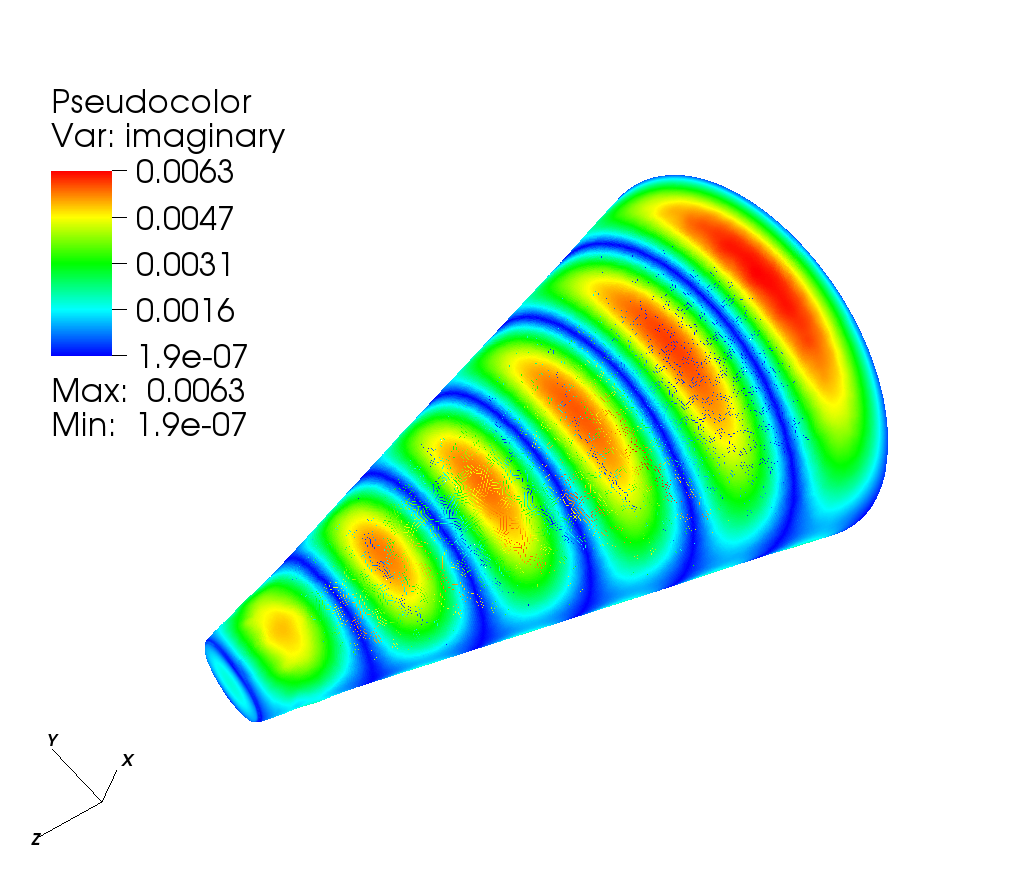}}%
      \caption{ Magnitude of surface current density on the truncated cone: (a) real part and (b) imaginary part }
	\label{fig:cone}
\end{figure}

\begin{figure}[!ht]
\centering
\includegraphics[width=0.8\linewidth]{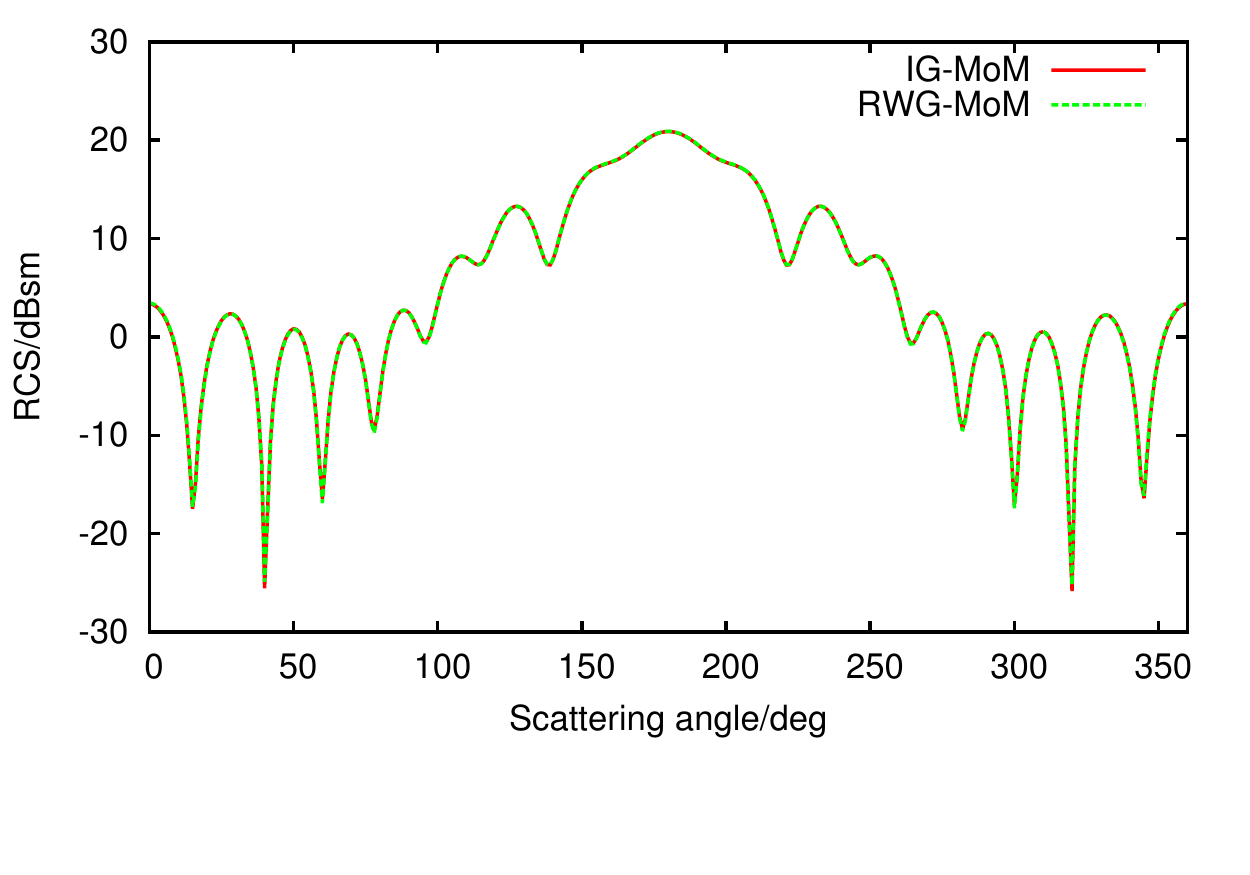}
\caption{Radar cross section of the truncated cone ($\phi=0$ cut)}
	\label{fig:cone_rcs}
\end{figure}

The first example is a truncated cone, with the height $3.0\lambda$ and the radii of the top and bottom circular cross-sections being $0.2\lambda$ and 1$\lambda$, respectively.  A plane wave that is polarized along $\hat{x}$ and propagating in the $-{\hat z}$ direction is incident on the object. The object is represented using 12378 nodes; this number is increased at the top and bottom surfaces so as to maintain a sufficient degree of sharpness. The number of DoFs for IGA-MoM and conventional MoM are $24756$ and $37128$, respectively. The surface current density obtained using IGA-MoM is depicted in Fig. \ref{fig:cone}. It is evident from these plots that the currents are smooth (without unphysical aberrations anywhere, especially near the crease). Fig. \ref{fig:cone_rcs} compares the radar cross section between the proposed method and the conventional method of moments, and it is apparent that the agreement between the two is excellent. 

\begin{figure}[!ht]
\centering
\subcaptionbox{\label{fig:cubecyl1_real}}
		[0.5\linewidth]{\includegraphics[width=0.45\linewidth]{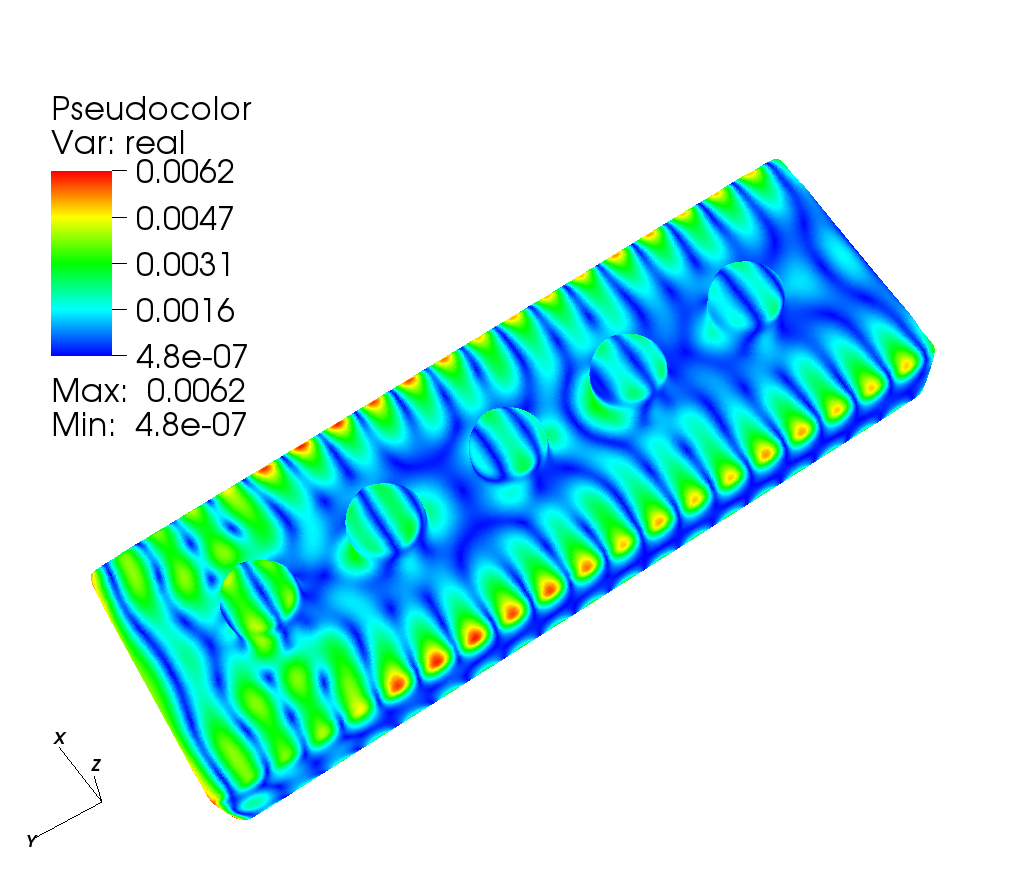}}%
      \subcaptionbox{\label{fig:cubecyl1_imag}}
		[0.5\linewidth]{\includegraphics[width=0.45\linewidth]{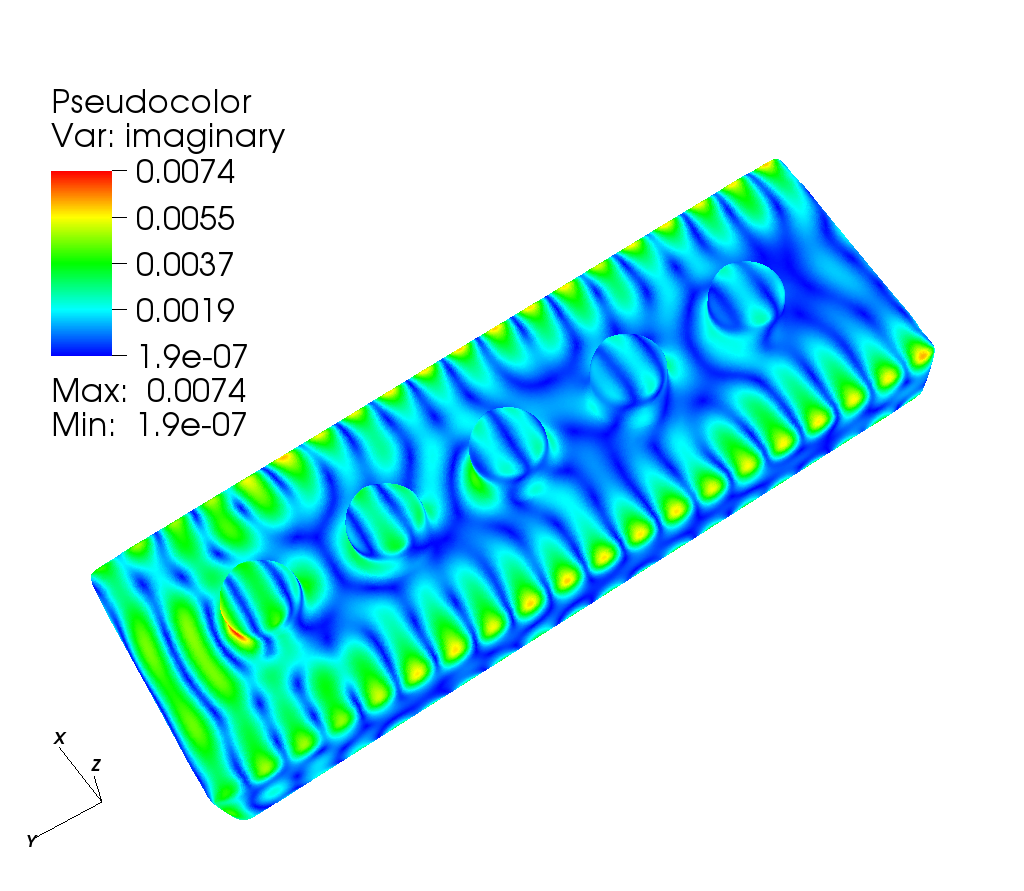}}%
      \caption{ Magnitude of surface current density on the composite structure: (a) real part and (b) imaginary part }
	\label{fig:cubecyl1}
\end{figure}

The second example is a structure composed of a block ($3.33\lambda \times 10.0\lambda \times 1.33\lambda $) and five cylinders (radius is $0.5\lambda$ and the height is $0.67\lambda$) uniformly distributed on the top surface of the block. An electromagnetic field that is propagating along  $-{\hat y}$ and polarized along the $\hat{z}$ incident on this object. The number of DoF for the IGA-MoM and conventional MoM are $29028$ and $43536$, respectively.   Figure\ref{fig:cubecyl1} presents the real and imaginary parts of the current distribution on the surface of the object, and again, it is evident that the results are smooth without artifacts. Further, excellent agreement can be seen between IGA-MoM and conventional MoM in the RCS data. 
\begin{figure}[!ht]
\centering
\includegraphics[width=0.8\linewidth]{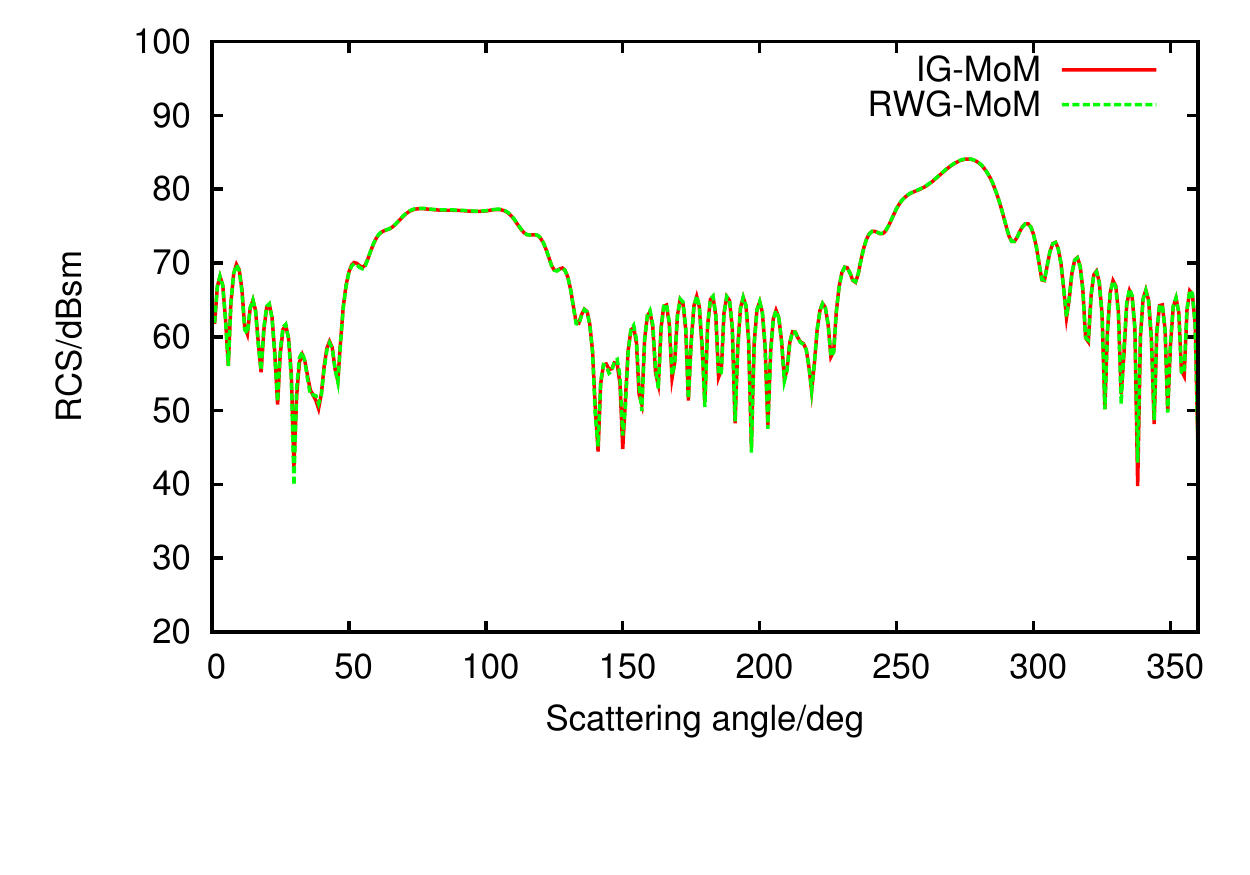}
\caption{Radar cross section of the composite structure ($\phi=90$ cut)}
	\label{fig:cubecyl_rcs}
\end{figure}

\begin{figure}[!ht]
\centering
\subcaptionbox{\label{fig:plane_real}}
		[0.5\linewidth]{\includegraphics[width=0.45\linewidth]{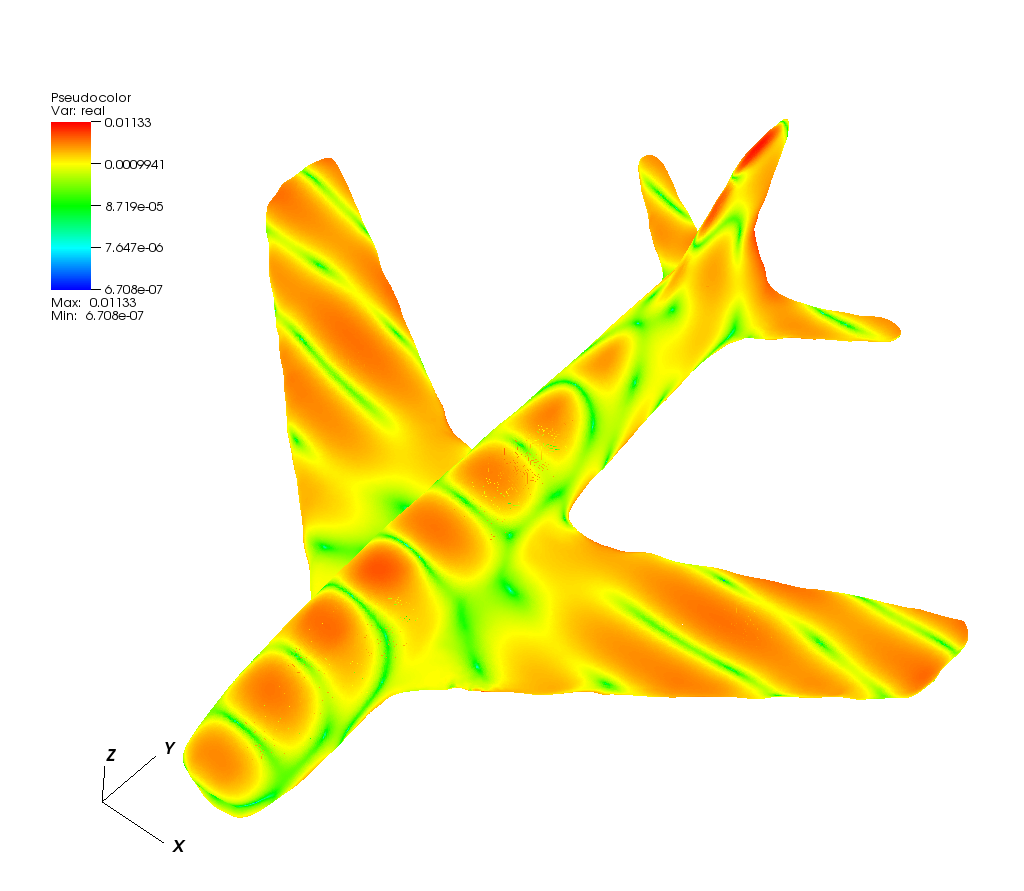}}%
      \subcaptionbox{\label{fig:plane_imag}}
		[0.5\linewidth]{\includegraphics[width=0.45\linewidth]{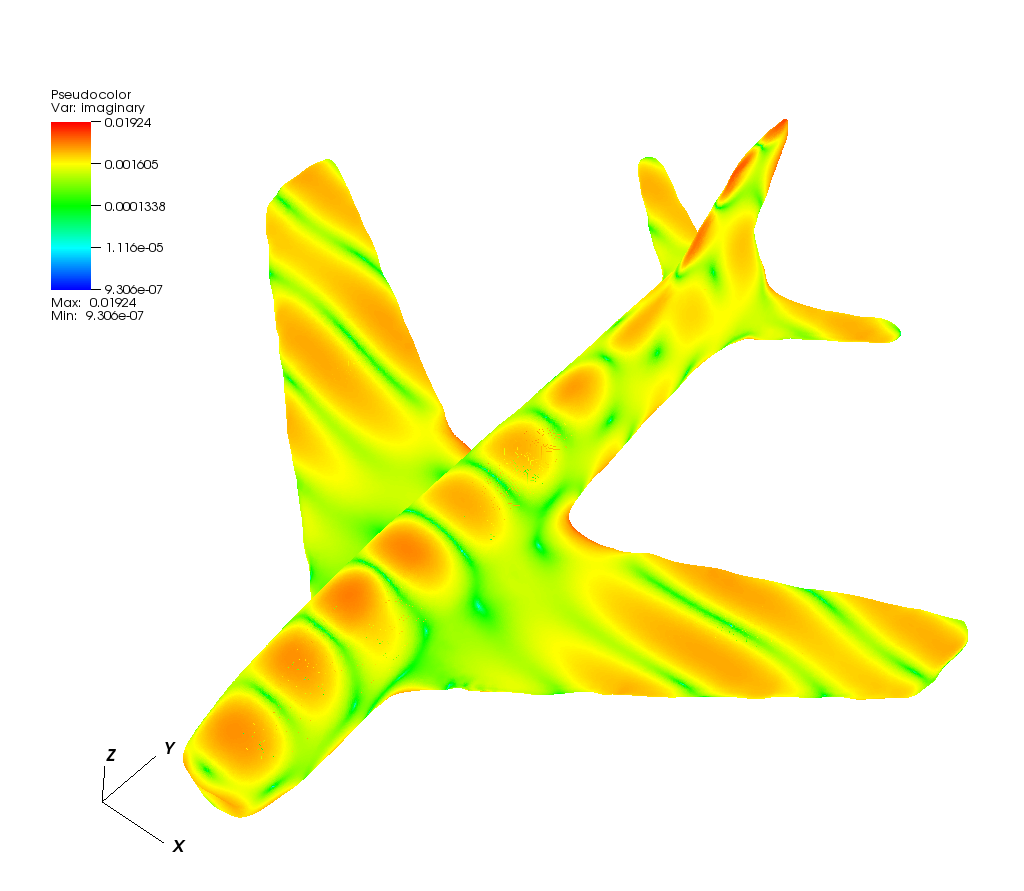}}%
      \caption{ Magnitude of surface current density on the air plane model: (a) real part and (b) imaginary part }
	\label{fig:plane}
\end{figure}

\begin{figure}[!ht]
\centering
\includegraphics[width=0.8\linewidth]{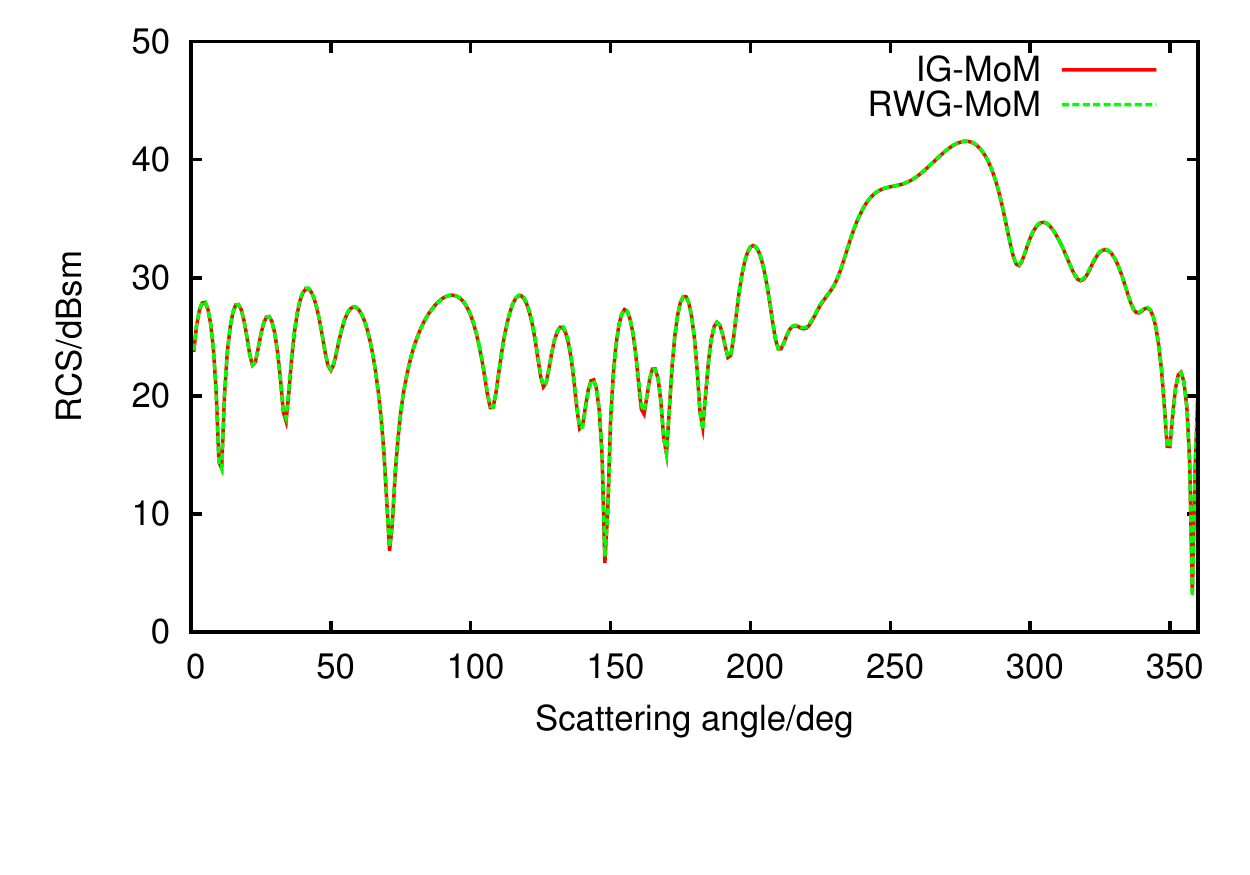}
\caption{Radar cross section of the air plane model ($\phi=90$ cut)}
	\label{fig:plane_rcs}
\end{figure}

As a final example in this set, we analyze scattering from a model airplane whose dimensions are  $5.49\lambda \times 5.48\lambda \times 1.52\lambda $. The plane wave incident on the object is propagating along   $-{\hat y}$ and polarized along $\hat{z}$. As before, data is obtained using both IGA-MoM and conventional MoM using $23988$ and $35976$ DoFs, respectively. The current distribution using IGA-MoM is depicted  in Fig.\ref{fig:plane}. Again, it is evident that the current is well captured as is the scattering cross-section (see Fig. \ref{fig:plane_rcs}). 

\subsection{Conditioning of Isogeometric System at Low-frequency}
Next, we discuss the stability of the resulting system (and its preconditioned form) at low frequency using IGA basis and the aforementioned wavenumber scaling. The metrics used for the discussion are both the condition number of the resulting system and the number of iterations necessary to achieve a desired error tolerance when solving this system using an iterative solver. In all the examples presented, we use a generalized minimal residual (GMRES) iterative solver and the object being analyzed  is a sphere with radius 1$m$ that is discretized such that the average distance between vertices is around 0.15$m$. 

\begin{figure}
\centering
\includegraphics[width=0.8\linewidth]{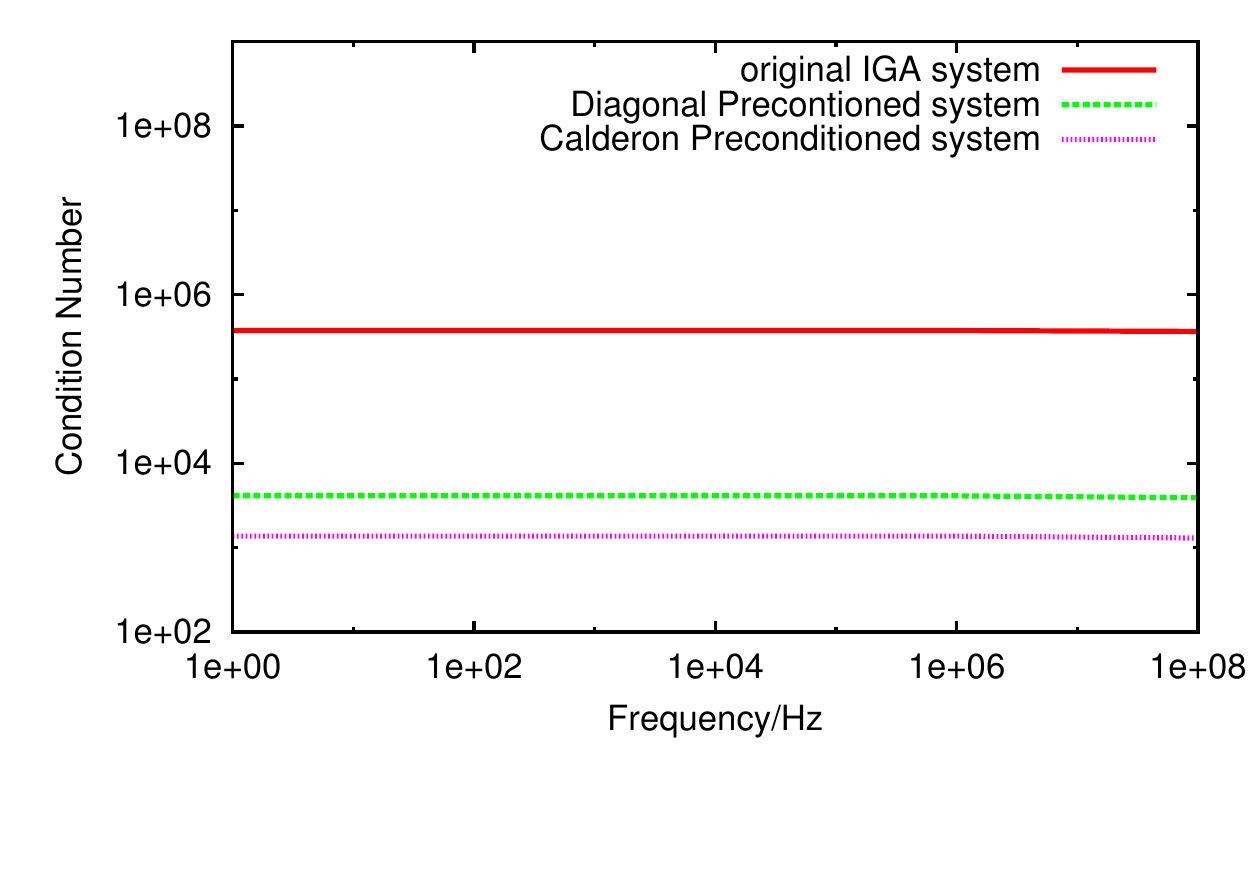}
\caption{Condition number at different frequencies}
	\label{fig:cond_no}
\end{figure}

Fig. \ref{fig:cond_no} depicts the condition number of IGA-MoM systems as a function of frequency from $1$Hz to $100$MHz. The three curves are, respectively, 
for the original IGA system, the diagonal preconditioned system, the Calderon preconditioned system. From the plots, it is evident that the condition number is almost constant over the whole band for all the  systems. Both diagonal preconditioner and application of the Calder\'{o}n precondition offer significant improvements over the original IGA-MoM. Note, the behavior of the original IGA-MoM is very much unlike conventional MoM, thanks to the properties of the basis function used. 

\begin{figure}
\centering
\includegraphics[width=0.8\linewidth]{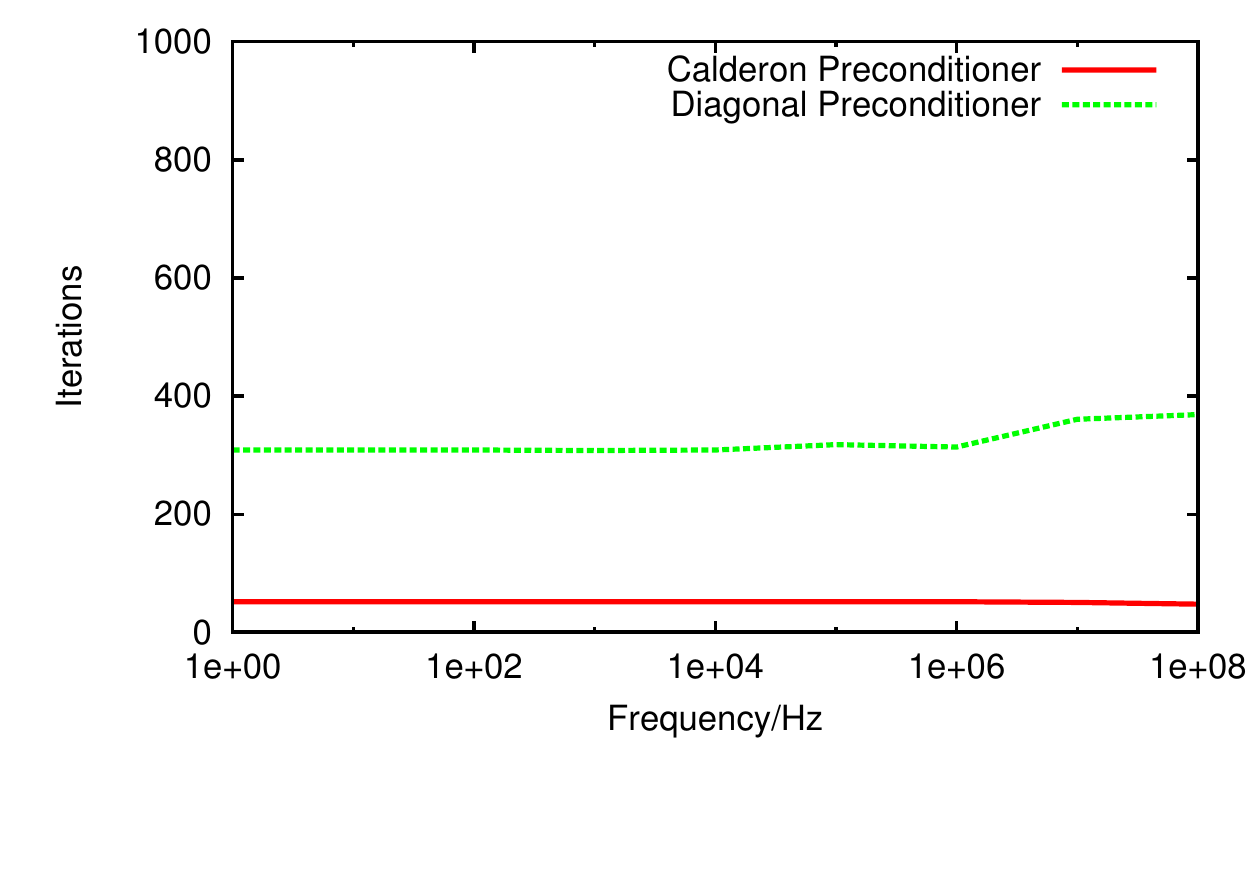}
      \caption{Number of GMRES iterations to converge to a tolerance of  $1.0\times10^{-6}$}
	\label{fig:iter_no1}
\end{figure}

\begin{figure}
\centering
\includegraphics[width=0.8\linewidth]{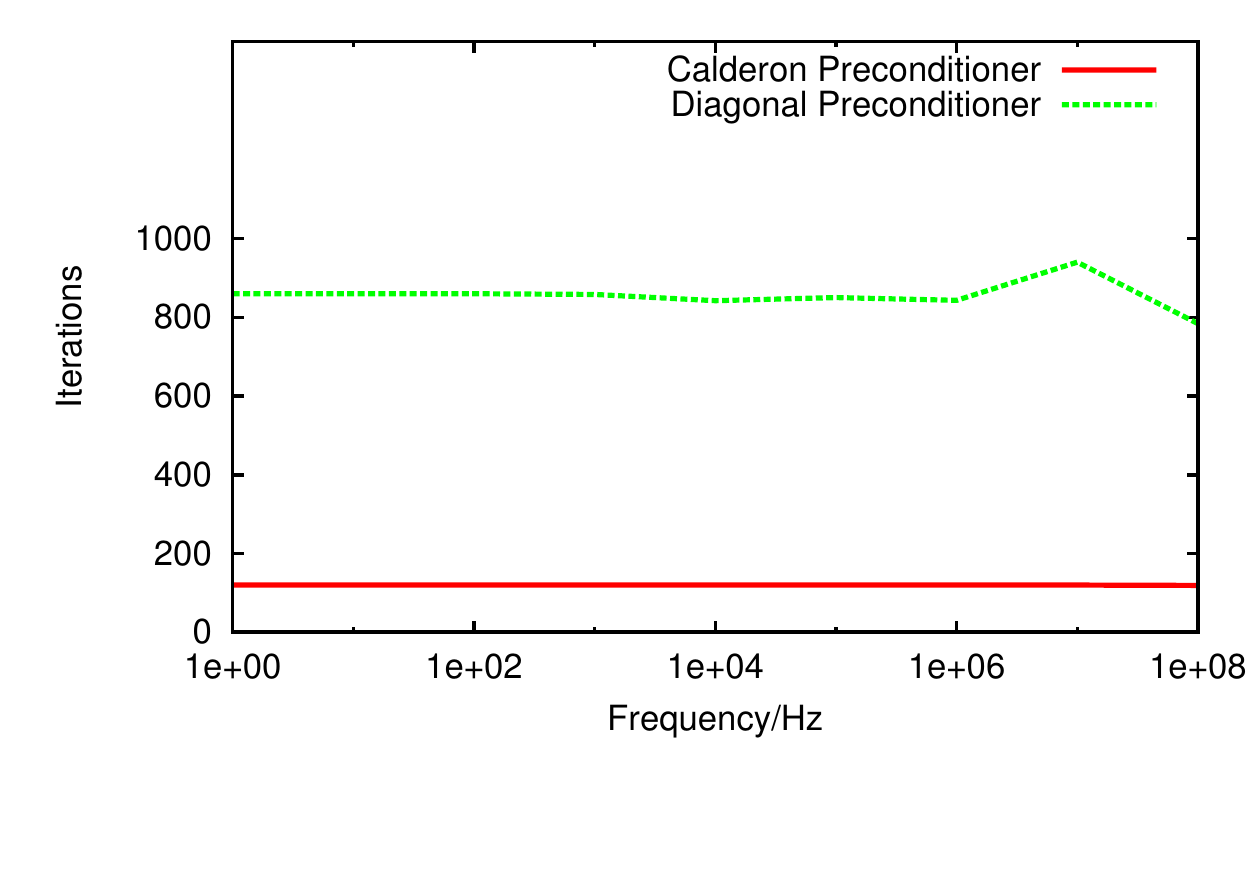}
      \caption{Number of GMRES iterations to converge to a tolerance of $1.0\times10^{-10}$}
	\label{fig:iter_no2}
\end{figure}

Next, we examine the number of GMRES iterations required to converge to an error tolerance of $1.0\times 10^{-6}$ and $1.0\times 10^{-10}$ at various frequencies. These results are depicted in Figs.  \ref{fig:iter_no1} and Fig. \ref{fig:iter_no2}, respectively, both sampled at $1$Hz, $100$Hz, $10^4$Hz, $10^6$Hz and $10^8$Hz. Since the iterative numbers for the original IGA system are much higher than that of  the diagonal preconditioned and Calderon preconditioned system, only the iteration numbers for the latter two systems are given. 
As seen in the two plots, the number of iterations required for both systems are relatively constant across a range of frequencies (from 1Hz to 100MHz). 

\begin{figure}
\centering
\includegraphics[width=0.8\linewidth]{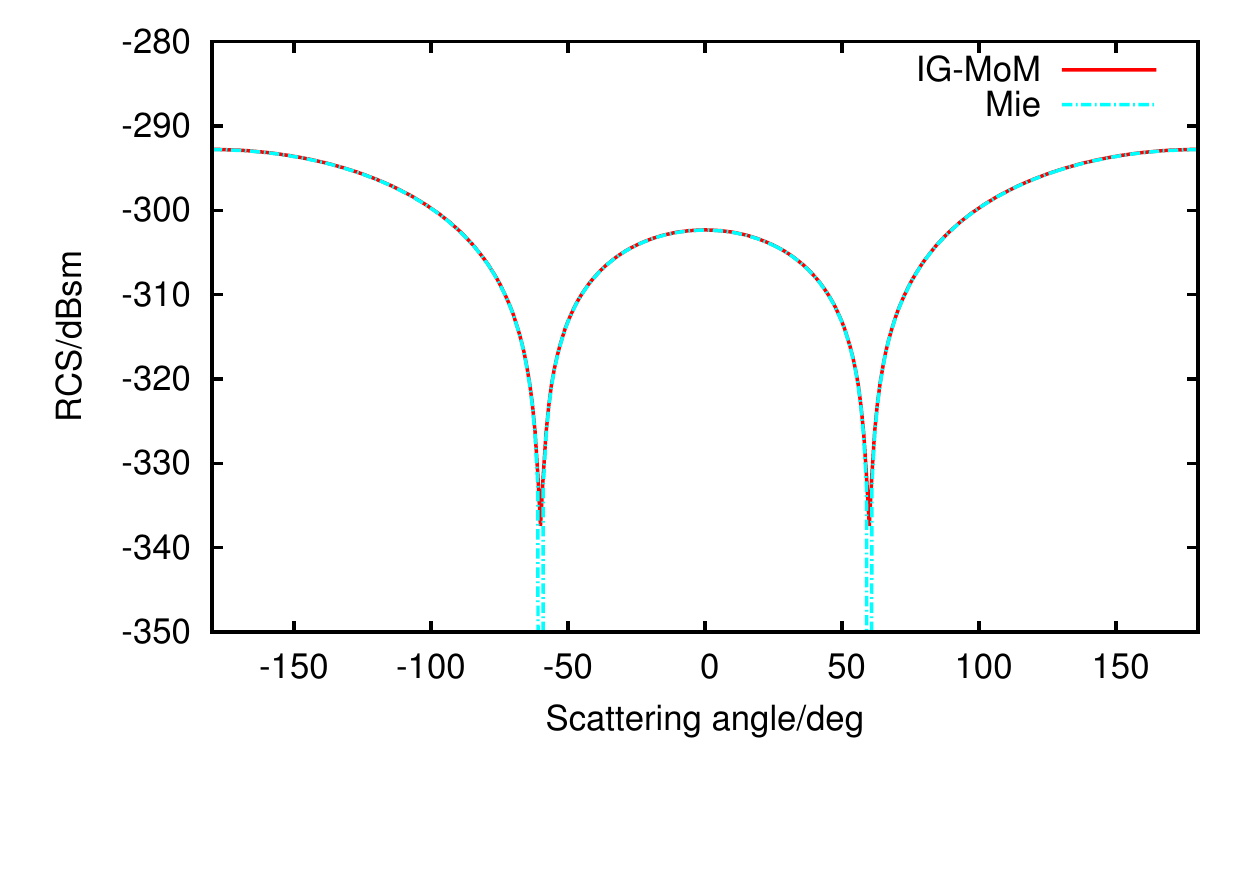}
      \caption{Comparison of RCS data obtained at 1Hz with analytical solutions}
	\label{fig:lowFreq}
\end{figure}

These two tests serve to illustrate some of the salient features of the proposed approach. (i) Low frequency stability is achieved simply by diagonal preconditioning. This is in contrast with conventional approaches where it is explicitly necessary to define auxiliary unknowns {\em albeit} at linear cost. (ii) Imposition of the Calder\'{o}n preconditioner involves an inversion of the Gramm matrix as an addition operation. The original matrix system is retained which makes integration with fast methods trivial. This is different from existing methods that require Buffa-Christiansen basis functions to be defined on  barycentric meshes. Finally, as is evident from Fig. \ref{fig:lowFreq}, the results obtained from the Calder\'{o}n preconditioned solver at 1Hz agrees very well with analytical data obtained. 

\subsection{Examples with Multi-scale mesh}
\begin{figure}[!ht]
\centering
\subcaptionbox{\label{fig:sph_ms}}
		[0.5\linewidth]{\includegraphics[width=0.45\linewidth]{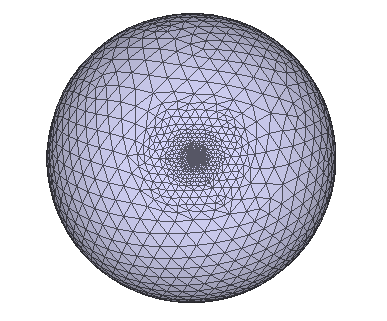}}%
      \subcaptionbox{\label{fig:sph_ms1}}
		[0.5\linewidth]{\includegraphics[width=0.45\linewidth]{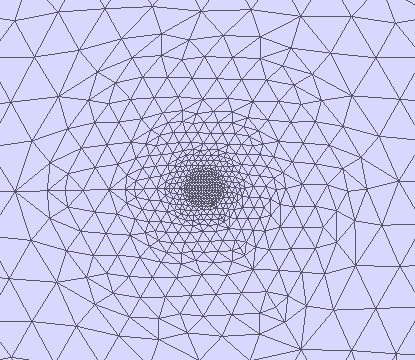}}%
      \caption{  Multiscale control mesh: (a) the whole mesh and (b) locally refined region}
	\label{fig:sph_ms_mesh}
\end{figure}

\begin{figure}[!ht]
\centering
\includegraphics[width=0.8\linewidth]{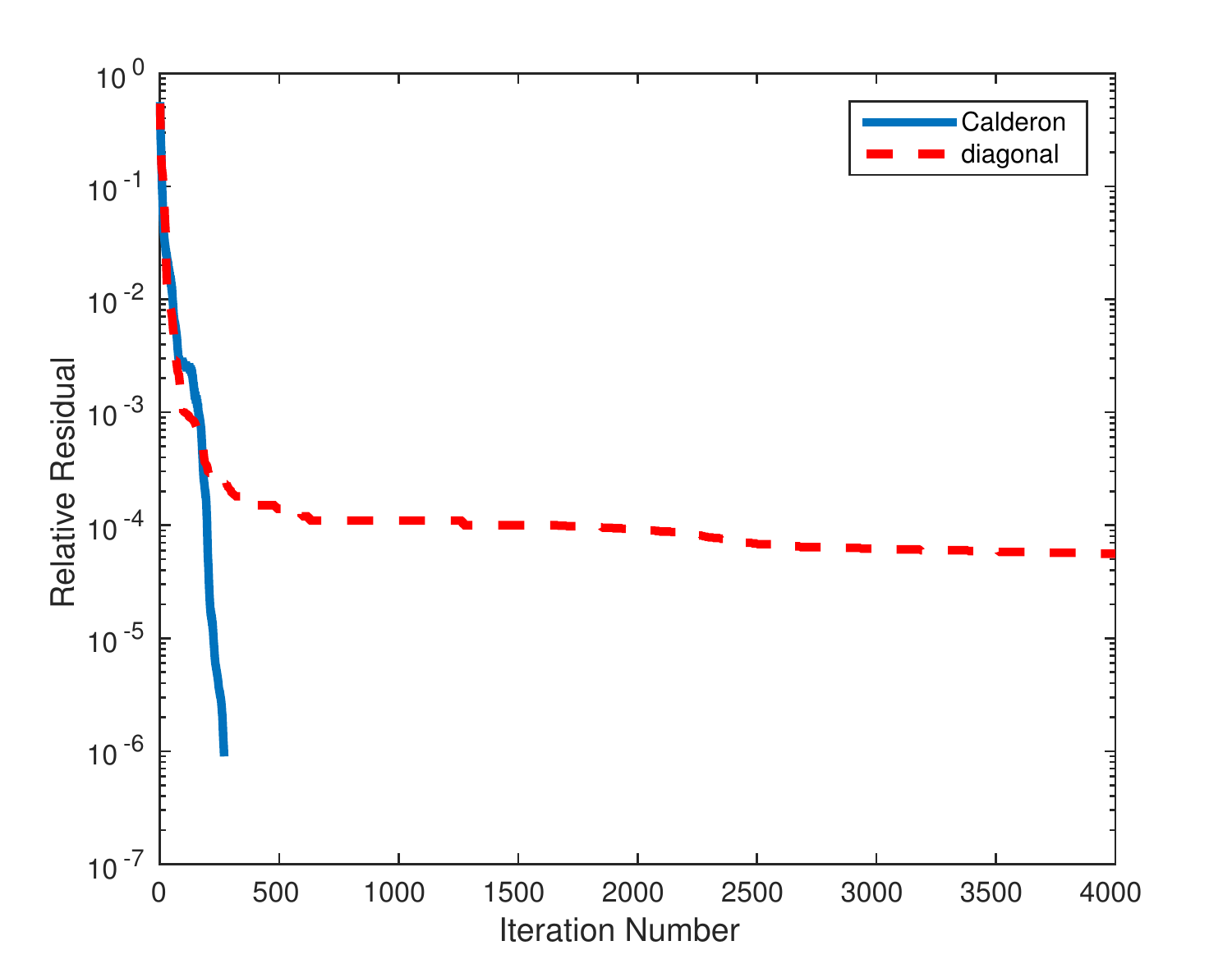}
\caption{Convergence of the iterative solver}
\label{fig:sph_ms_iter}
\end{figure}

In all the examples analyzed thus far, the initial control mesh is such that the resulting tesselation is almost uniform in that the ratio of the maximum edge length to the minimum edge length is ${\cal O}(1)$. However, a more intellectually interesting and practically applicable problem is when this ratio is significantly higher. It is challenging to design stable methods for these problems due to two effects that act in concert with each other; wavenumber and element size scaling. In this example, scattering from a sphere with locally refined meshes is simulated, and verified by comparing with the one with almost uniform
initial control mesh. Fig. \ref{fig:sph_ms_mesh} demonstrates the multiscale control mesh for the unit sphere. The radius of the sphere is 1$\lambda$, and the incident plane wave is polarized along the $\hat{x}$ direction and propagating in the $-\hat{z}$ direction. The resulting system of equations is solved using GMRES. Figure \ref{fig:sph_ms_iter} compares the convergence history for both the Calder\'{o}n preconditioned system as well as a diagonal preconditioned system. The efficacy of the Calder\'{o}n preconditioned system is evident by the rapid convergence. 

\section{Summary} \label{sec:summary}
In this paper, we have developed a novel isogeometric analysis technique for solving the electric field integral equation that is encountered in electromagnetic field analysis. The fundamental thesis of isogeometric methods is that they inherit the rather significant advantages of modern CAD tools--smoothness of geometry representation, morphing, dynamic meshes (if necessary), etc--by using the same basis sets for both geometry representation as well as representation of physics on this geometry. The choice of the electric field integral equation is predicated upon the fact that it is one of the more challenging equations to solve. The approach presented here leverages existing geometric construction techniques that use subdivision to define basis functions that result in well behaved integral operators. Thanks to these operators, it is possible to trivially modify these to impose Calder\'{o}n preconditioners, and construct systems that are low frequency stable and can handle multiscale geometric features. 
A number of results demonstrate the convergence and accuracy of the technique, applicability to computation of scattering at both regular and low frequencies, as well as to structures with multiscale features. It is very simple to apply the proposed basis set to other types of integral equations, such as magnetic field integral equation and combined field integral equation. However, as with the introduction of any new technique several open problems remain: these include addition of features to handle edges and open domains, integration with fast solvers, development of techniques for multiply connected objects, using this framework to include Debye sources, etc. Several of these topics are active areas of research within the group and will be presented in different forums soon.   

\section*{Acknowledgments}
This work was supported in part by NSF CMMI-1250261. The authors thanks the high performance computing center at Michigan State University for computational support. 

\section*{References}

\bibliography{iga}

\end{document}